\documentclass[%
reprint,onecolumn,
amsmath,amssymb,
aps,
pra,
longbibliography,
]{revtex4-1}

\usepackage{graphicx}
\usepackage{bm}
\usepackage{amsmath}
\usepackage{float}
\usepackage{natbib}

\usepackage[whole]{bxcjkjatype}

\newcommand{\rp}{r_0^{(p)}}

\begin{document}

\preprint{APS/123-QED}

\title{Non-Kolmogorov scaling
	for two-particle relative velocity\\
	in two-dimensional inverse energy-cascade turbulence
}

\author{Tatsuro Kishi} 
 \email{tatsuro@kyoryu.scphys.kyoto-u.ac.jp}
\author{Takeshi Matsumoto}
\author{Sadayoshi Toh}
\affiliation{
	Division of Physics and Astronomy, 
	Graduate School of Science, 
	Kyoto University,
	Kitashirakawa Oiwaketyo Sakyoku, Kyoto 606-8502, Japan
}%
\date{\today}

\begin{abstract}
Herein, we numerically examine the relative dispersion of Lagrangian particle pairs
in two-dimensional inverse energy-cascade turbulence.
Behind the Richardson--Obukhov $t^3$ law of relative separation, 
we discover that
the second-order moment of the relative velocity 	
have a temporal scaling exponent different from 
the prediction based on the Kolmogorov's phenomenology.
The results also indicate that time evolution of the probability distribution function 
of the relative velocity is self-similar. 
The findings are obtained by enforcing the Richardson--Obukhov law 
either by considering a special initial separation or by conditional sampling. 
In particular, we demonstrate that the conditional sampling
removes the initial-separation dependence
of the statistics of the separation and relative velocity.
Furthermore, we demonstrate that 
the conditional statistics are robust with respect to the change in the parameters 
involved, and that the number of the removed pairs from the sampling 
decreases when 
the Reynolds number increases. 
We also discuss the insights gained as a result of conditional sampling.
\end{abstract}

\maketitle

\section{Introduction\label{sec:introduction}}
Relative dispersion has been widely investigated 
following a pioneering study by Richardson \cite{LewisFRichardson1926},
who observed the super-diffusive manner of separation 
between two particles in the atmosphere.
The study introduced the diffusion-type differential equation
for the probability distribution function (PDF) of separation, $r$.
A significant part of this equation is that it includes the diffusion 
coefficient dependent on $r$ itself.
Furthermore, Richardson predicted the celebrated $t^3$ law
of the second-order moment of the relative separation from the PDF,
$\langle r^2(t) \rangle \propto t^3$, 
and this is referred to as the Richardson--Obukhov law.
The scaling argument leading to this law (which was developed first for the three 
dimensional (3D) turbulence) can be applicable to two-dimensional (2D) turbulence, 
as reviewed in \cite{Salazar2008}.
In this study, we restrict our attention to 
the relative dispersion in 2D turbulence. 

The $t^3$ prediction was performed prior to Kolmogorov's phenomenology
for 3D turbulence proposed in 1941 (K41) \cite{Kolmogorov1941}, and 
later demonstrated as consistent with the K41 dimensional analysis
\cite{Obukhov1941,Batchelor1950a}.
With respect to the 2D turbulence, specifically, in the inverse energy-cascade state,
the Richardson--Obukhov law was similarly derived from a 2D analog of the K41,
which was developed by Kraichnan, Leith, and Batchelor \cite{Kraichnan1967,Leith1968,Batchelor1969}; hereinafter, the analog is referred to as K41 for convenience.
Specifically, based on the phenomenologies, 
the second-order moment of the relative separation 
in the inertial range can assume the following form:
\begin{align}
	\langle r^2(t) \rangle \simeq 
	\begin{cases}
		\langle r_0^2 \rangle + S_2(r_0)t^2 & (t \ll t_{\mathrm{B}}),\\
		g \varepsilon  t^3 & (t_\mathrm{B} \ll t \ll T_L),
	\end{cases}
	\label{eq:Batchelor}
\end{align}
where $r_0 \equiv |\bm{r}_0|$ 
denotes the initial separation of the pairs,
$\varepsilon$ denotes the energy dissipation rate or
the energy flux in the inertial range,
$S_2(r) = C_2 \varepsilon^{2/3}r^{2/3}$ 
denotes the second-order longitudinal velocity structure function, 
$C_2$ is a constant, 
$t_\mathrm{B}=r_0^{2/3} \varepsilon^{-1/3}$ 
denotes the Batchelor time,
$T_L$ denotes the integral time scale,
and $g$ denotes the Richardson constant.
Up to the Batchelor time $t_\mathrm{B}$, 
each particle moves with the initial velocity. 
Subsequently, the relative separation becomes independent of $r_0$
and behaves according to the $t^3$ law, exhibiting super-diffusivity (Richardson--Obukhov regime). 

With respect to 2D inverse energy-cascade turbulence, the $t^3$ law
is observed 
in laboratory experiments \cite{Jullien1999,VonKameke}
for appropriately selected initial separations.
Recently, Rivera and Ecke \cite{Rivera2005,Rivera2016} 
performed experiments
by varying initial separations and observed
that the power-law exponent of $\langle r^2(t) \rangle$ 
in the inertial range  depends on the initial separation $r_0$.
They also observed $t^3$-scaling behavior similar to
the Richardson--Obukhov scaling law 
only for a certain range of initial separations.
The initial separation dependence and 
existence of special initial separations leading to the $t^3$ law
were observed in 2D direct numerical simulation (DNS) 
\cite{Boffetta2002} as well. 
With respect to the 3D direct energy-cascade turbulence, 
the situation is similar:
the slope of $\langle r^2(t) \rangle$ as a function of $t$ 
varies due to the length of initial separations
in laboratory experiments \cite{Ott2000}.
Recently, DNSs in 3D also indicated that 
the $t^3$ law only appears for a certain selected initial separation
\cite{Yeung2004,Sawford2008,Bitane2013}.

Based on 2D and 3D results, 
the conclusion at currently achievable Reynolds numbers
is that the time evolution of $\langle r^2(t) \rangle$ 
strongly depends on the initial separation.
Thus, the Richardson--Obukhov $t^3$ law emerges 
only for a selected initial separation, 
and this is termed as \textit{the proper initial separation} 
in the current study (as detailed in Sec.~\ref{sec:pis}).
The problem to be solved is the dependence of the $t^3$ law on the initial separation; specifically, whether the $t^3$ law observed for the special initial separation is relevant with the K41 or just coincidental. 
It is known that the initial separation dependence is alleviated by considering $\langle |\bm{r}-\bm{r_0}|^2 \rangle$ instead of $\langle r^2 \rangle$.
 By analyzing $\langle |\bm{r}-\bm{r_0}|^2 \rangle$ 
at sufficiently high Reynolds numbers, 
Bitane et al. \cite{Bitane2012,Bitane2013} introduced the modified scaling law including a subleading term,
$\langle | \bm{r}(t) - \bm{r}_0 |^2 \rangle = g\varepsilon t^3 (1 + C t_0/t)$ 
for $t \gg t_0$, where $t_0$ denotes a time scale of convergence to
Richardson--Obukhov regime, $t_0 = S_2(r_0) / 2 \varepsilon$, and
$C$ denotes a parameter based on $r_0$.
It is noted that $C=0$ for $r_0 = 4 \eta$, where
$\eta$ denotes the Kolmogorov length scale.
The $r_0 = 4 \eta$ is termed as "optimal choice" in their study and can correspond to the proper initial separation.
Furthermore, Buaria et al. \cite{D.Buaria2015} suggested 
an asymptotic state, and this is independent of the initial separations.
The same authors \cite{Buaria2016} investigated turbulent relative dispersion 
utilizing diffusing/Brownian particles, i.e., particles of various Schmidt numbers ($Sc$)
with white/Brownian noise added to their trajectories. 
They found that the initial separation dependence is weaker and Richardson scaling 
is more robust  for $Sc = O(1)$ than $Sc =\infty$ (fluid particles).

Several studies \cite{Boffetta2002,Rivera2005}
in the 2D inverse-energy cascade turbulence
discussed the proper initial separation,
and concluded that the $t^3$-scaling behavior observed only for the special initial separation 
is an artifact caused by the finite-size effect of the limited inertial range. 
Given the aforementioned reasons, 
they argued that proper initial separation exists 
even in the low Reynolds-number simulations and that 
the proper initial separation is significantly lower 
than the smallest lower bound of the inertial range.
In particular, the observed scaling law
$\langle r(t)^2 \rangle \propto t^3$ started
to hold outside of the inertial range,
as already noted in \cite{Kellay2002}. 
Subsequently, the $t^3$ law with the proper initial separation
extends into the inertial range. 
However, details of the finite-size effects, e.g.,
the dependence of the $t^3$ law on the width of the inertial range, 
remains to be clarified.

There is another problem with respect to the proper initial separation.
The K41 can be applied to the two-particle Lagrangian 
relative velocity, and predicts $t^1$
scaling for the second-order moment as 
\begin{equation}
	\langle v^2(t)\rangle \propto \varepsilon t^1,
	\label{eq:two_particle_law}
\end{equation}
in the inertial range, where $v(t)$ denotes the relative velocity.
In recent 3D numerical studies \cite{Yeung2004,Bitane2013},
the relative velocity is also observed to depend on the initial separation
such as relative separation.
Furthermore, it appears that the second-order moment of 
the relative velocity exhibits a different scaling exponent 
from the K41 prediction.

The long-standing problem of two-particle relative diffusion
in turbulence can be the applicability of the K41 to the Lagrangian relative separation
and velocity statistics and at least at presently available Reynolds numbers.
It is well-known that the K41 scaling does not precisely hold,  particularly for the 3D turbulence
due to the intermittency effect. However, the deviation from the K41 is small with respect to the low-order statistics of the Eulerian velocity 
such as the energy spectrum or the second-order structure functions. Thus, 
the K41 is successful for the Eulerian velocity. In contrast,
the K41 appears to fail in describing the second-order moments of
the relative separation and velocity, which are Lagrangian quantities, to the same extent as the low-order 
Eulerian velocity. This large gap between Eulerian and Lagrangian statistics should be filled.
It is possible that the gap is caused by a finite Reynolds-number effect. 

In this study, 
we numerically examine two-particle relative diffusion 
in 2D energy inverse-cascade turbulence with either normal viscosity or hyperviscosity.
The main reason for selecting the 2D system is that
detailed numerical studies (e.g., a large number of particle-pair samples and 
long-time integration) are more feasible. 
Furthermore, the Eulerian velocity is 
intermittency free \cite{Paret1997,Boffetta2000}, and consequently, corresponds to ``an ideal framework
to examine Richardson scaling in Kolmogorov turbulence'',
as noted by \cite{Boffetta2002}.
Thus, we can factor out the intermittency effect on
the deviation of the Lagrangian statistics from the K41 prediction 
when we analyze 2D results.
Evidently, limitations exist while selecting the 2D system.
As aforementioned, there are common problems 
in the Richardson--Obukhov law in 2D and 3D systems. 
However, their nature is not necessarily identical.
Careful discussion and further investigations are required
while applying our results in this study to the 3D case. 
Nevertheless, insights obtained here in 2D 
can be useful in addressing the 3D problem.

We conduct our numerical study as follows.
First, we develop a conditional sampling to remove the initial-separation dependence.
We demonstrate that the conditioned $\langle r^2(t) \rangle$ 
curves of various initial separations collapse on 
the unconditioned curve starting from the proper initial separation. 
From the robustness, we infer that
the $t^3$ law of the proper initial separation is 
consistent with the K41.
We then discuss the generality of the conditional sampling,
namely, the dependence of the conditioned results
on the parameters of the conditional sampling.
Finally, we examine the scaling behavior of the relative velocity
with and without the conditional sampling in detail.

The two main results obtained in 2D energy inverse-cascade turbulence are:
(i) relative velocity deviates from the K41 scaling ,i.e., scaling law (\ref{eq:two_particle_law}),
although the relative separation obeys the Richardson--Obukhov $t^3$ law; 
and (ii) relative velocity is self-similar (intermittency free).

Both suggest that the K41 does not hold for second-order statistics of relative velocity.

Sec.~\ref{sec:dns} 
presents the details of our 2D numerical study.
Sec.~\ref{sec:pis} 
introduces a working hypothesis and
describes the proper initial separation.
In Sec.~\ref{sec:sampling},
we describe our conditional sampling and
discuss what can be inferred from conditional statistics on the relative separation.
Sec.~\ref{sec:results}
presents statistics on the relative velocity 
with and without conditional sampling.

\section{
Numerical simulation method
\label{sec:dns}
}
\begin{table}
	\begin{center}
	\begin{tabular*}{\textwidth}{@{\extracolsep{\fill} } cccccccccccccc}\\
		$N^2$  & $\delta x$ & $\delta t$ & $\nu$ & $h$ &$\alpha$ & $k_f$ & $\varepsilon_{in}$ & $\varepsilon$ & $\sigma_\varepsilon$ & $L$ & $u_{\mathrm{rms}}$ & $Re_\alpha$ & $N_p^2$\\
		$1024^2$  & $0.006$ & $0.002$ & $1.8 \times 10^{-38}$ & 8 & $35$ & $249$ & $0.1$ & $0.019$ & $2.9 \times 10^{-4}$ & $0.38$ & $0.5$ & $40$ & $2048^2$\\
		$2048^2$ & $0.003$ & $0.001$ & $4.664 \times 10^{-43}$ & 8 & $35$ & $496$ & $0.1$ & $0.019$ & $2.9 \times 10^{-4}$ & $0.37$ & $0.5$ & $80$ & $2048^2$\\
		$4096^2$ & $0.0015$ & $0.001$ & $1.13 \times 10^{-47}$ & 8 & $35$ & $997$ & $0.1$ & $0.018$ & $2.6 \times 10^{-4}$ & $0.36$ & $0.5$ & $160$ & $2048^2$\\
		$2048^2$ & $0.003$ & $0.004$ & $7.666\times 10^{-6}$ & 1 & $3.005$ & $200$ & $3.027\times 10^{-4}$ & $5.28\times 10^{-5}$ & $3.35\times 10^{-5}$ & $0.47$ & $0.076$ & $39$ & $2048^2$
	\end{tabular*}
	\caption{Parameters of numerical simulations: 
		$N^2$, $\delta x = 2\pi / N$, $\delta t$, 
		$\nu$, $h$, $\alpha$, $k_f$, $\varepsilon_{in}$,
		$\varepsilon$, $\sigma_\varepsilon$, 
		$L$, $u_{\mathrm{rms}}$, $Re_\alpha$ and $N_p^2$ 
		denote the number of grid points, grid spacing, 
		size of the time step, 
		 (hyper)viscosity coefficient,
	        order of the Laplacian of the (hyper)viscosity,
		hypodrag coefficient, 
		forcing wavenumber, 
		energy input rate of the forcing, 
		mean of the resultant energy flux in the inertial range,  
		standard deviation of the resultant energy flux,
		integral scale,  
		root-mean-square velocity,
		infrared Reynolds number and  
		number of the Lagrangian particles, 
		respectively.}
	\label{tab:parameter}
	\end{center}
	\hrulefill
\end{table}

We mainly consider pair-dispersion statistics 
in a statistically steady,
homogeneous, and isotropic 2D inverse-energy
cascade turbulent velocity field $\bm{u}(\bm{x},t)$.
In the velocity field, we perform a set of DNSs of  
the 2D incompressible Navier-Stokes equation
in a doubly periodic square of side length, $2\pi$.
We integrate the equation in the form of vorticity, $\omega(\bm{x}, t) = 
\partial_x u_y({\bm x}, t) - \partial_y u_x({\bm x},t)$, which is
\begin{equation}\label{eq:NS}
	\frac{\partial \omega}{\partial t} + 
	\left(\bm{u}\cdot\nabla\right)\omega = 
	(-1)^{h + 1}\nu\Delta^h\omega + \alpha\Delta^{-1}\omega + f.
\end{equation}
The setting and our numerical method are identical to those used 
in \cite{XIAO2009,Mizuta2013}.
Here, $\nu$ denotes the (hyper)viscosity coefficient
and $\alpha$ denotes the hypodrag coefficient.
The order of the Laplacian of the (hyper)viscosity, $h$, is
set to $8$ or $1$.
The forcing term, $f({\bm x}, t)$, is 
given in terms of the Fourier coefficients,
$\hat{f}(\bm{k},t) = k^2\varepsilon_{in}
/[n_f \hat{\omega}^{*}(\bm{k},t)]$, 
where 
\, $\hat{f}$ \, denotes the Fourier transform of the function 
$f(\bm{x}, t)$.
The energy input rate is denoted by $\varepsilon_{in}$, and
$n_f$ denotes the number of the Fourier modes 
in the following forcing wavenumber range.
We select the coefficients, $\hat{f}({\bm k}, t)$, 
as non-zero only in high wave numbers, $\bm{k}$,
satisfying $k_f - 1 < |\bm{k}| < k_f + 1$. 
Thus, the energy input rate is maintained as constant in time.
Numerical integration of Eq.~(\ref{eq:NS}) 
is performed via the pseudospectral method 
with the 2/3 dealiasing rule in space and 
the 4-th order Runge--Kutta method in time.
Table \ref{tab:parameter} lists the parameters of 
simulations used in the study.

\begin{figure}
	\centerline{\includegraphics[keepaspectratio,scale=0.55]{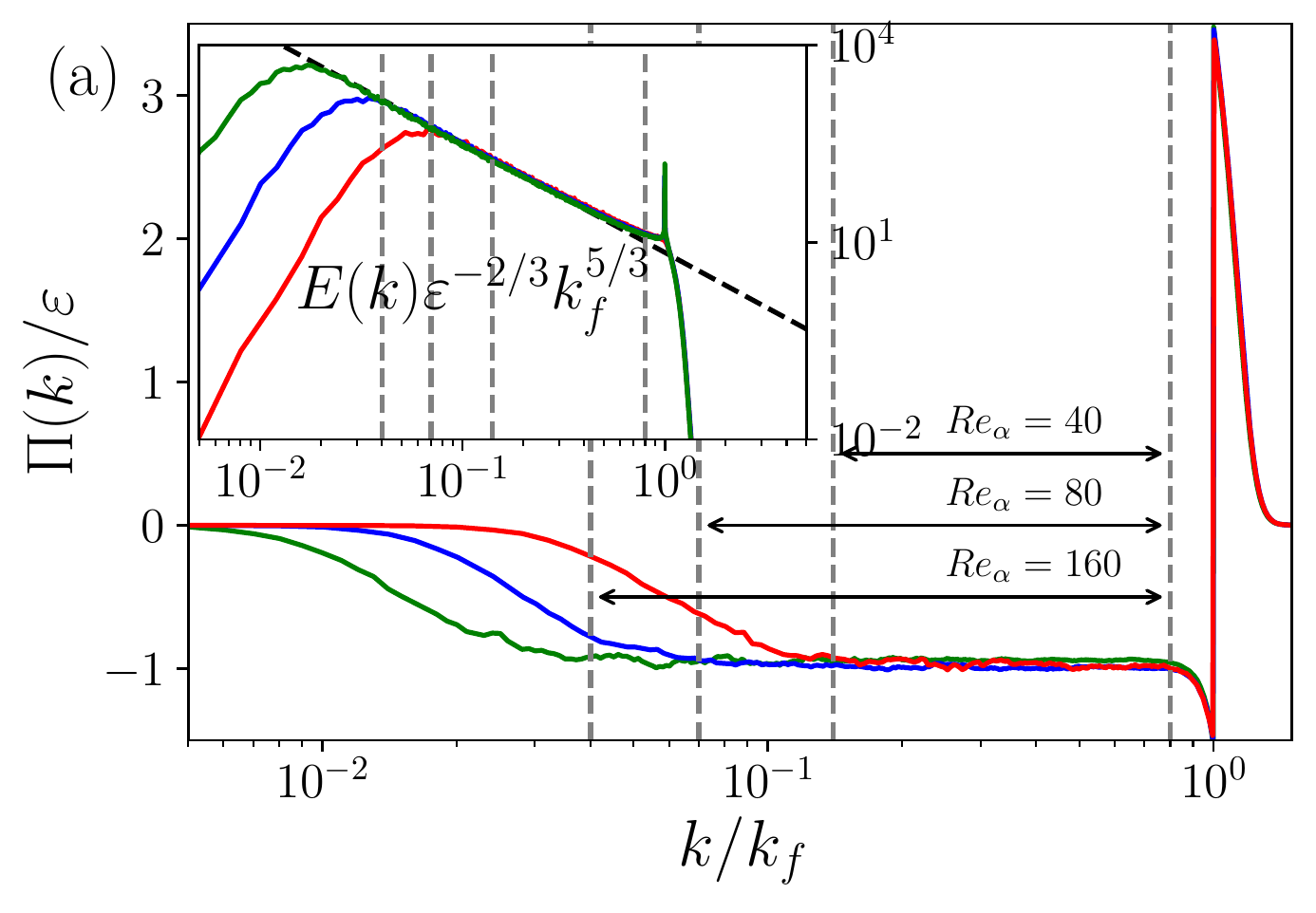}
		\includegraphics[keepaspectratio,scale=0.55]{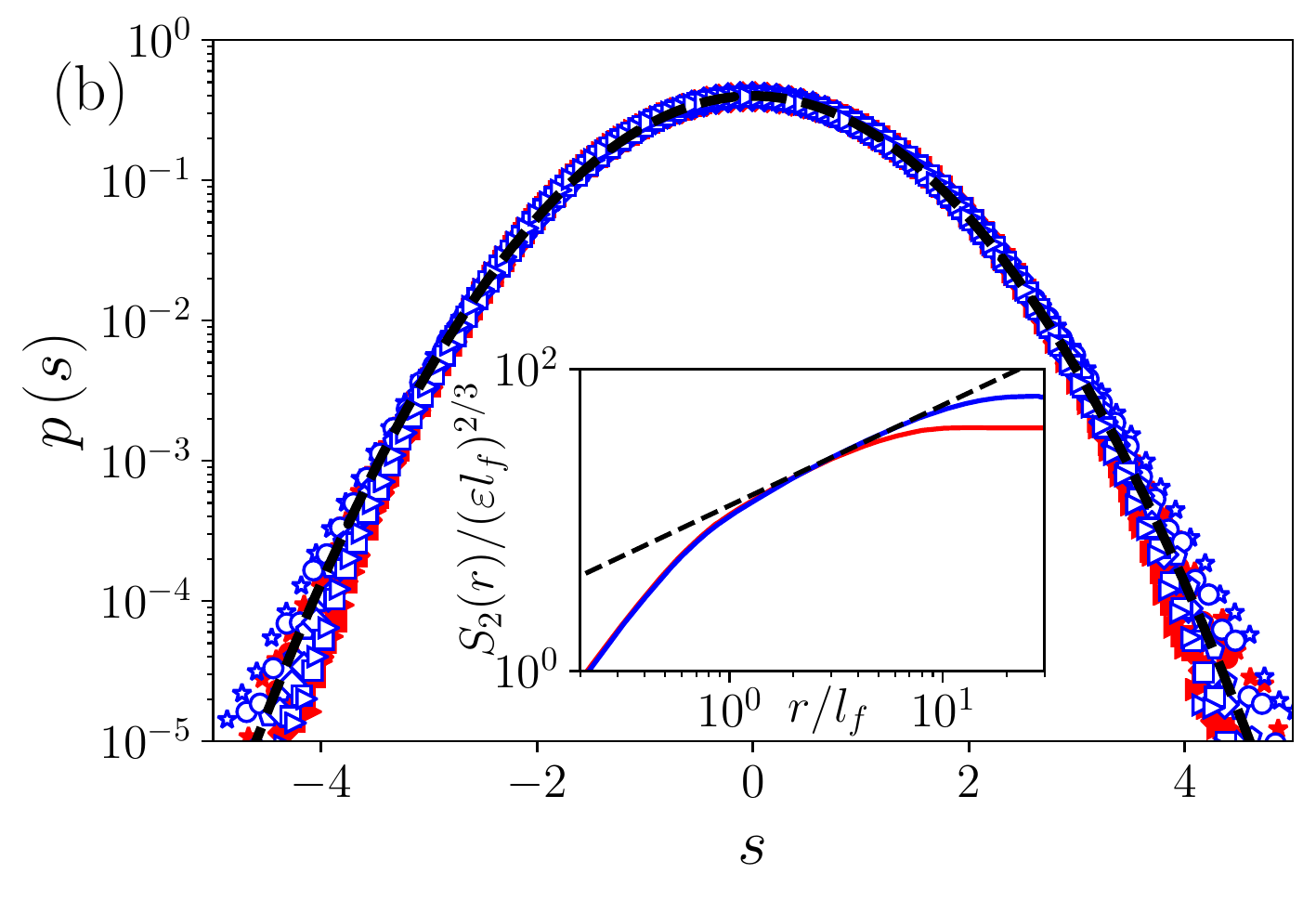}}
	\centerline{\includegraphics[keepaspectratio,scale=0.55]{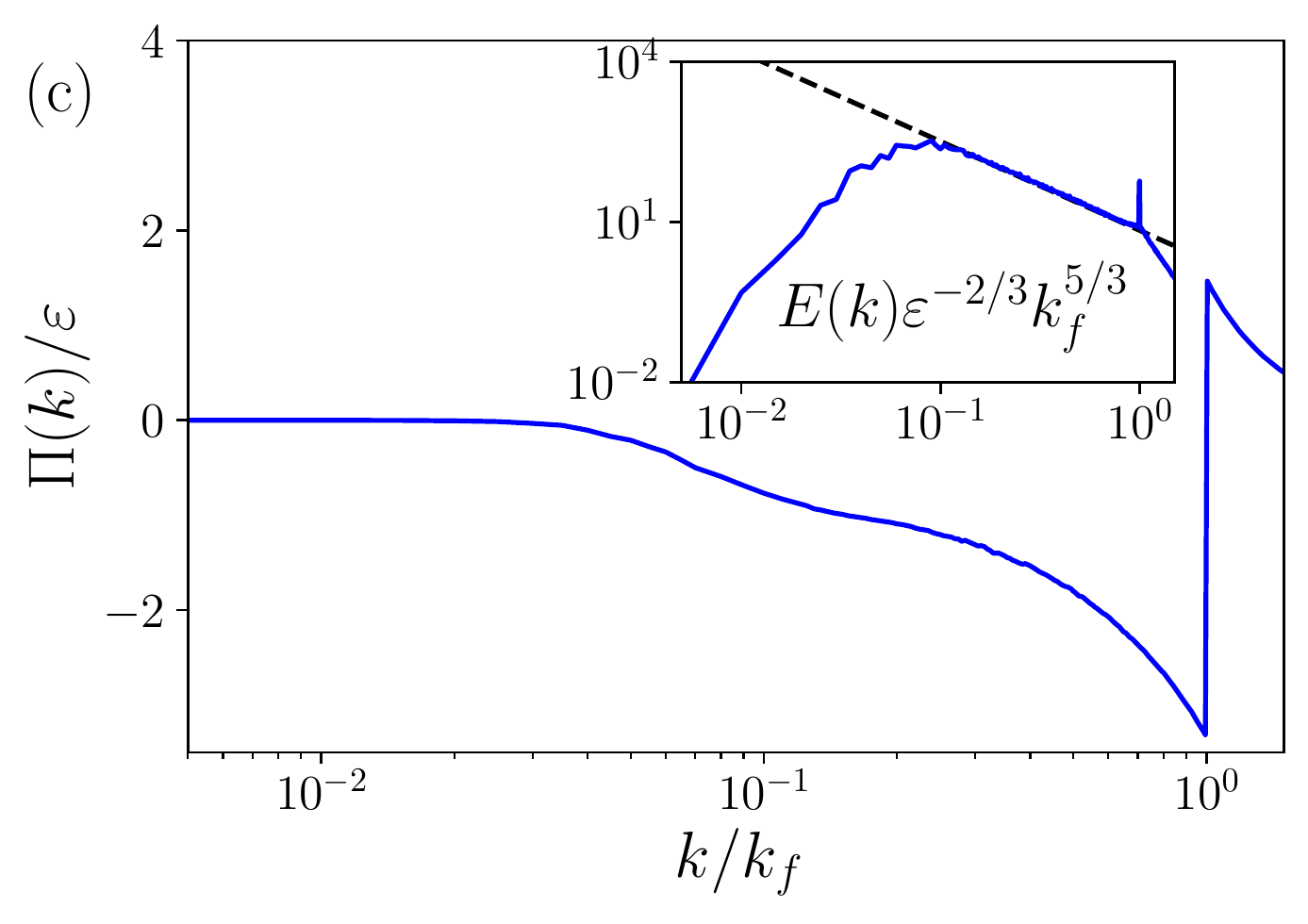}
 		\includegraphics[keepaspectratio,scale=0.55]{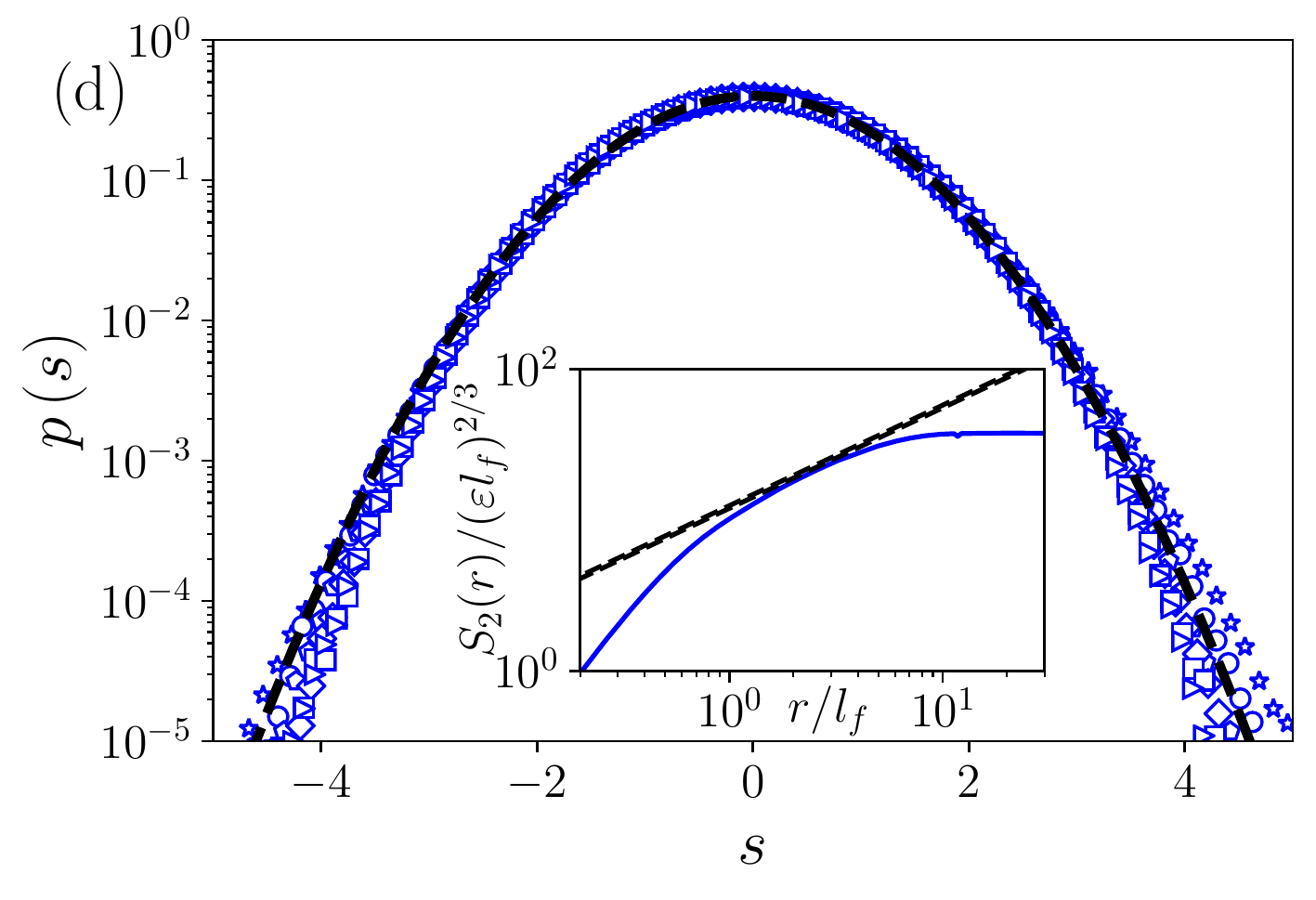}}
	\caption{
		(a) Time-averaged energy flux $\Pi(k)$ 
		for $Re_\alpha = 40$ (red line),
		$Re_\alpha = 80$ (blue line) and $Re_\alpha = 160$ (green line).
		Ranges between the two gray dotted lines 
		correspond to the inertial ranges 
		as determined by the plateau regions of 
		the energy flux in the Fourier space.
		The inset shows 
		the time-averaged energy spectrum $E(k)$ 
		for $Re_\alpha = 40$ (red line),
		$Re_\alpha = 80$ (blue line) and 
		$Re_\alpha = 160$ (green line).
		(b) Normalized PDFs 
		of the longitudinal velocity increments 
		for $Re_\alpha=40$ (red) and $Re_\alpha = 80$ (blue)
		at various separations,
		$r/l_f = 1.1, 1.5, 2.2, 3.2, 4.8$ and $7.1$. 
                 Here, $l_f = 2\pi/k_f$ is the forcing scale and
                longitudinal velocity increment, $\delta u_l$, normalized by the second-order moment and
                is denoted by $s$: $s=\delta u_l/\sqrt{\langle \delta u_l^2 \rangle}$.
		The dotted line denotes the Gaussian distribution 
		with zero mean and unit variance. 
		The inset shows the second-order longitudinal structure function,
		$S_2(r)$
		for $Re_\alpha=40$(red) and $Re_\alpha = 80$(blue). 
		The dashed line represents the K41 scaling,
		$r^{2/3}$, for $S_2(r)$.
 		(c) Same as (a) albeit for the normal viscous case ($h = 1$).
		(d) Same as (b) albeit for the normal viscous case ($h = 1$).
 }
	\label{graph:euler}
\end{figure}

In the 2D energy inverse-cascade turbulence, the energy pumped in at the small scale
is transported to larger scales with a constant flux on average in the inertial range.
To measure this flux, we use the standard method to calculate the energy flux function
in the Fourier space. 
As shown in Fig.~\ref{graph:euler}(a) and (c), the flux
becomes wavenumber independent in the intermediate wavenumbers.
We consider the range of the wavenumbers as the inertial range.
Strictly speaking, a flat region is absent in Fig.~\ref{graph:euler} (c)
due to normal viscosity.
The energy flux in the inertial range
is equal to the energy dissipation rate taken out 
by the large-scale hypodrag,
$\varepsilon = \int_0^{\infty} 2 \alpha k^{-2}E(k) \mathrm{d}k$,
where $E(k)$ denotes the time-averaged energy spectrum. This corresponds to a standard method to numerically 
realize a statistically steady state of 2D energy inverse-cascade turbulence in a periodic domain.
A statistically steady state is judged from behavior of energy as a function 
of time. The typical wavenumber of the hypodrag is dimensionally estimated as
$(\alpha^3/\varepsilon)^{1/8}$, which is termed as the frictional wave number, $k_\alpha$.
Here, we use the infrared Reynolds number, $Re_\alpha \equiv k_f/k_\alpha$,
as proposed by Vallgren \cite{Vallgren2011} 
in order to quantify the span of the inertial range.
At the end of Sec.~\ref{sec:results}, we simulate a statistically 
quasi-steady state \cite{Kraichnan1967} by solving Eq.~(\ref{eq:NS})
without the hypodrag.

Subsequently, we demonstrate that the Eulerian statistics on the velocity field
are consistent with the established picture of the 2D inverse
energy-cascade turbulence.
As shown in the inset of Fig.~\ref{graph:euler}(a) and (c),
the energy spectra in the inertial range 
is consistent with the K41 and more precisely with 
the Kraichnan-Leith-Batchelor phenomenology.
Figures \ref{graph:euler}(b) and (d) show that
the PDFs of the longitudinal velocity increments,
$\delta u_l(r,t) = [\bm{u}(\bm{x}+\bm{r},t)-\bm{u}(\bm{x},t)]\cdot \bm{r}/r$,
at various $r$'s
in the inertial range collapse well to 
the Gaussian distribution irrespective 
of $r$, and this is in agreement with \cite{Boffetta2000}.
Here, $l_f = 2\pi/k_f$ denotes the forcing length scale.

To obtain the Lagrangian statistics, 
we employ a standard particle tracking method. 
The flow is seeded with a 
large number of tracer particles, i.e., $N_p^2$, in the velocity field.
The particles are tracked in time 
via integrating the advection equation,
\begin{equation}\label{eq:particle}
	\frac{\mathrm{d}}{\mathrm{d} t}\bm{x}_p(t) 
	= \bm{u}(\bm{x}_p(t),t),
\end{equation}
where $\bm{x}_p(t)$ denotes the particle position vector.
The numerical integration of Eq.~(\ref{eq:particle}) 
is performed using the Euler method. 
The velocity value at an off-grid particle position is
estimated by the fourth-order Lagrangian interpolation 
of the velocity calculated on the grid points.

The relative separation, $\bm{r}(t)$, is defined by 
$\bm{r}(t) = \bm{x}_1(t) - \bm{x}_2(t)$,
where $\bm{x}_1$ and $\bm{x}_2$ denote the positions of a particle pair.
The particles are initially seeded on square grid points 
where the grid spacing corresponds to $r_0$.
The statistics on the relative separation are calculated 
for the nearest neighbor particles at the initial time.
In this study, we vary the initial separation $r_0$ 
while maintaining the same total number, $N_p^2$, of the particles for each $r_0$.
For small values of $r_0$ (which are typically lower 
than the Eulerian grid size $\delta x$),
the initial particles do not cover the whole periodic domain.
We verify that the inhomogeneity of the initial positions 
of the particles does not affect Lagrangian statistics 
that are examined here
by comparing results
with different initial particle positions covering different parts 
of the periodic domain.
In addition to the separation, $r(t)$, 
our focus is on the longitudinal relative velocity 
of particle pairs as defined by
$v_l(t) \equiv [\bm{u}(\bm{x}_1(t),t)-\bm{u}(\bm{x}_2(t),t)]\cdot \bm{r}(t)/r(t)$.

We next discuss on how long we track the particles. We continue the tracking until
all the particle pairs leave the inertial range. 
We observe that this time typically concerns
$10$ large-scale eddy turn-over times ($L / u_{\rm rms}$) for the hyperviscous $Re_\alpha = 40$ case
and approximately $20$ turn-over times for the hyperviscous $Re_\alpha = 80$ case. With respect to each $r_0$, 
we perform the simulation of the duration twice.

The largest resolution simulation ($N^2 = 4098^2$) as listed in Table \ref{tab:parameter} 
is used only for confirming self-similarity of PDF of $v_l(t)$
in Sec.~\ref{sec:results}.
We define the Lagrangian average $\langle \cdot \rangle$ as 
$\langle A \rangle = \frac{1}{N_{\mathrm{adj}}}\sum_{i=1}^{N_{\mathrm{adj}}} A_i$,
where $A$ denotes any Lagrangian quantity and $A_i$ denotes a realization of $A$ 
by the $i$-th particle pair. 
$N_{\mathrm{adj}} = 2N_p(N_p-1)$ denotes the number of pairs of particles which adjoin each other at the initial time.

In the following sections,
we mainly use hyperviscosity rather than normal viscosity for DNSs.
This is because 
the hyperviscosity extends the inertial range for a given spatial
resolution. 
However, it is known to affect
the statistics at the transition
between the inertial and dissipation ranges \cite{Boffetta2012}.
Thus, it is possible that the hyperviscosity affects
particle-pair statistics.
Therefore, we perform hyperviscous 
and normal-viscous simulations
and confirm that the hyperviscosity does not affect the particle-pair
statistics.

\section{
Initial separation dependence of relative diffusion statistics
and conditional sampling
}

\subsection{Proper initial separation\label{sec:pis}}
\begin{figure}
	\centerline{
	\includegraphics[keepaspectratio,scale=0.55]{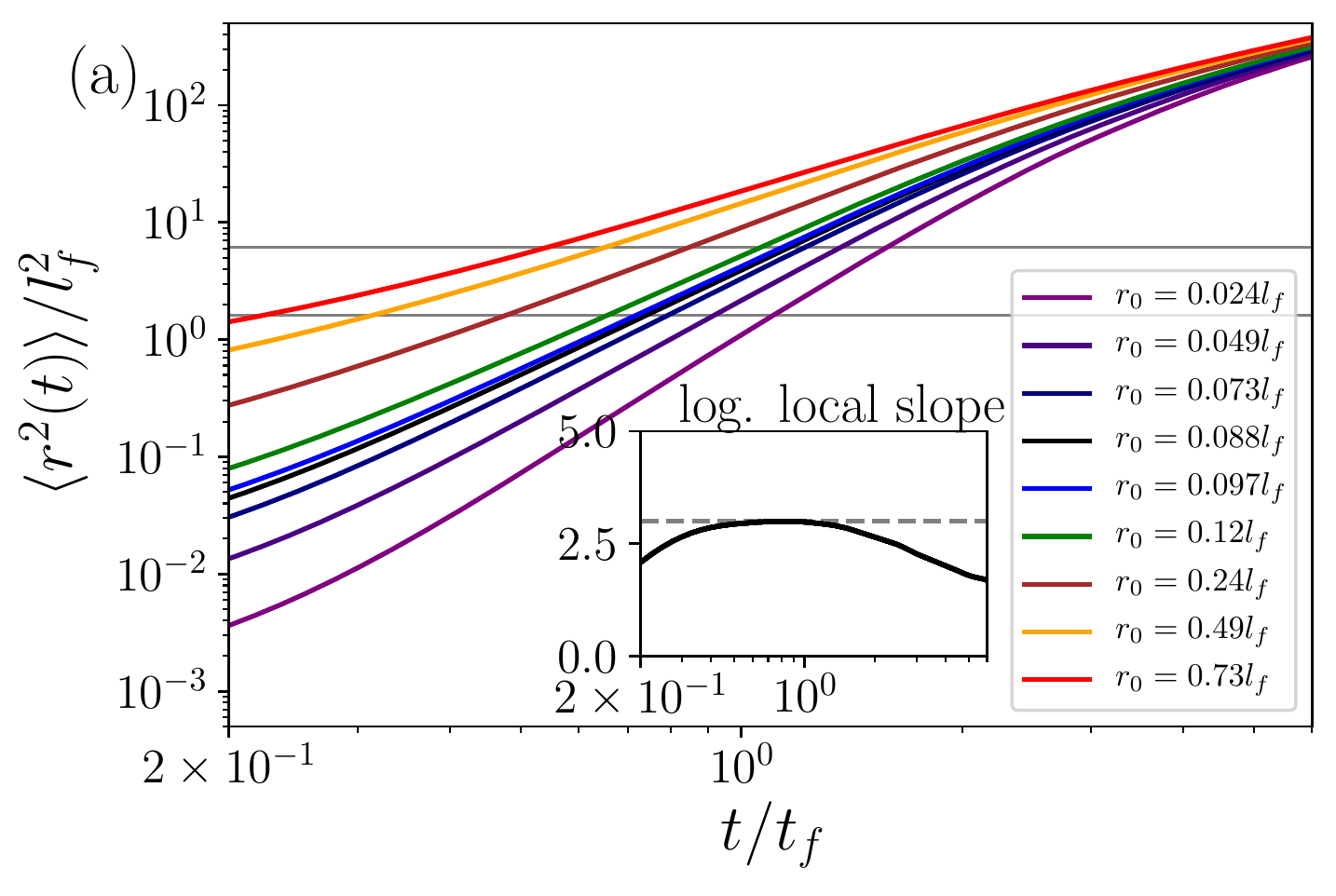}
	\includegraphics[keepaspectratio,scale=0.55]{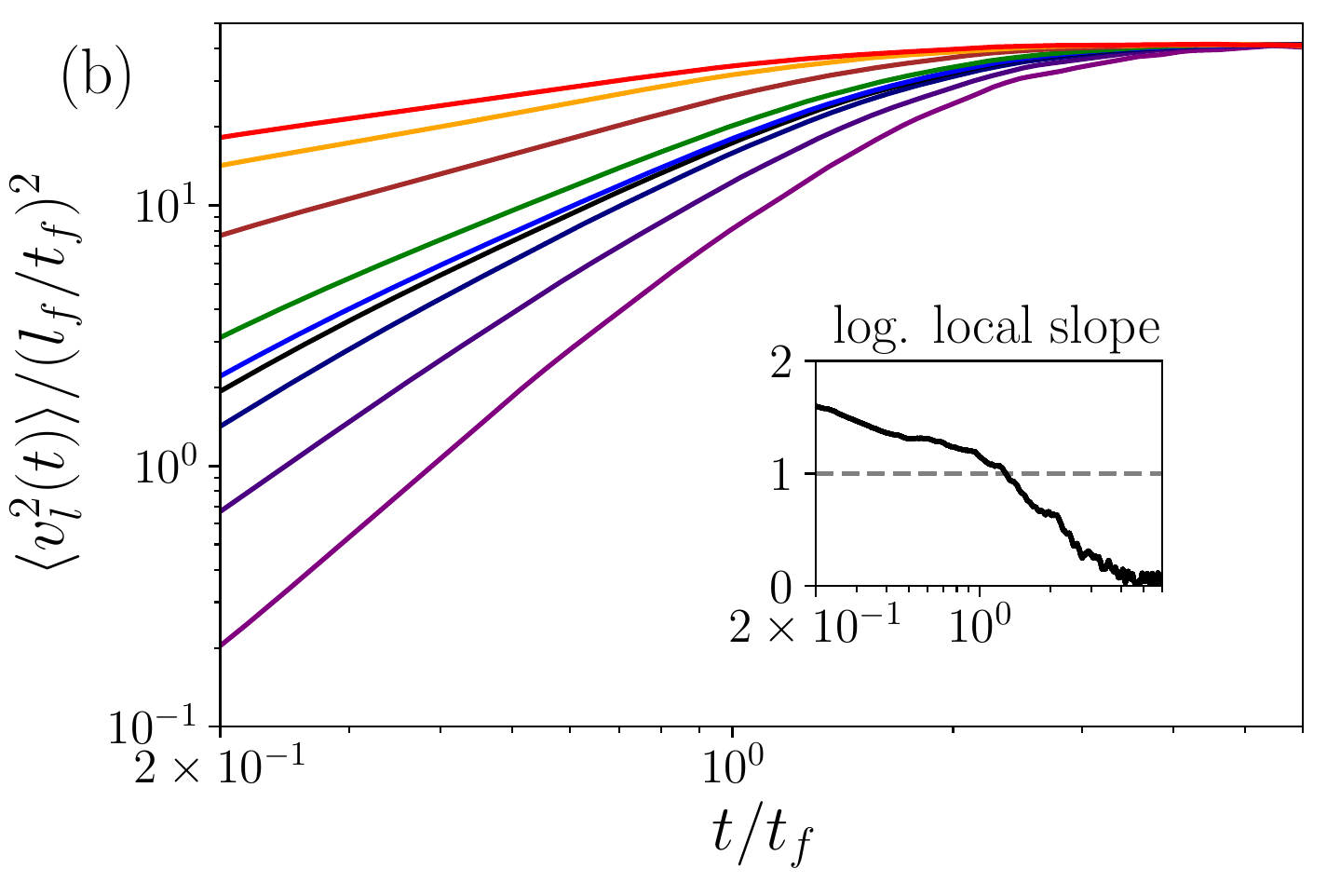}
	}
	\centerline{
	\includegraphics[keepaspectratio,scale=0.55]{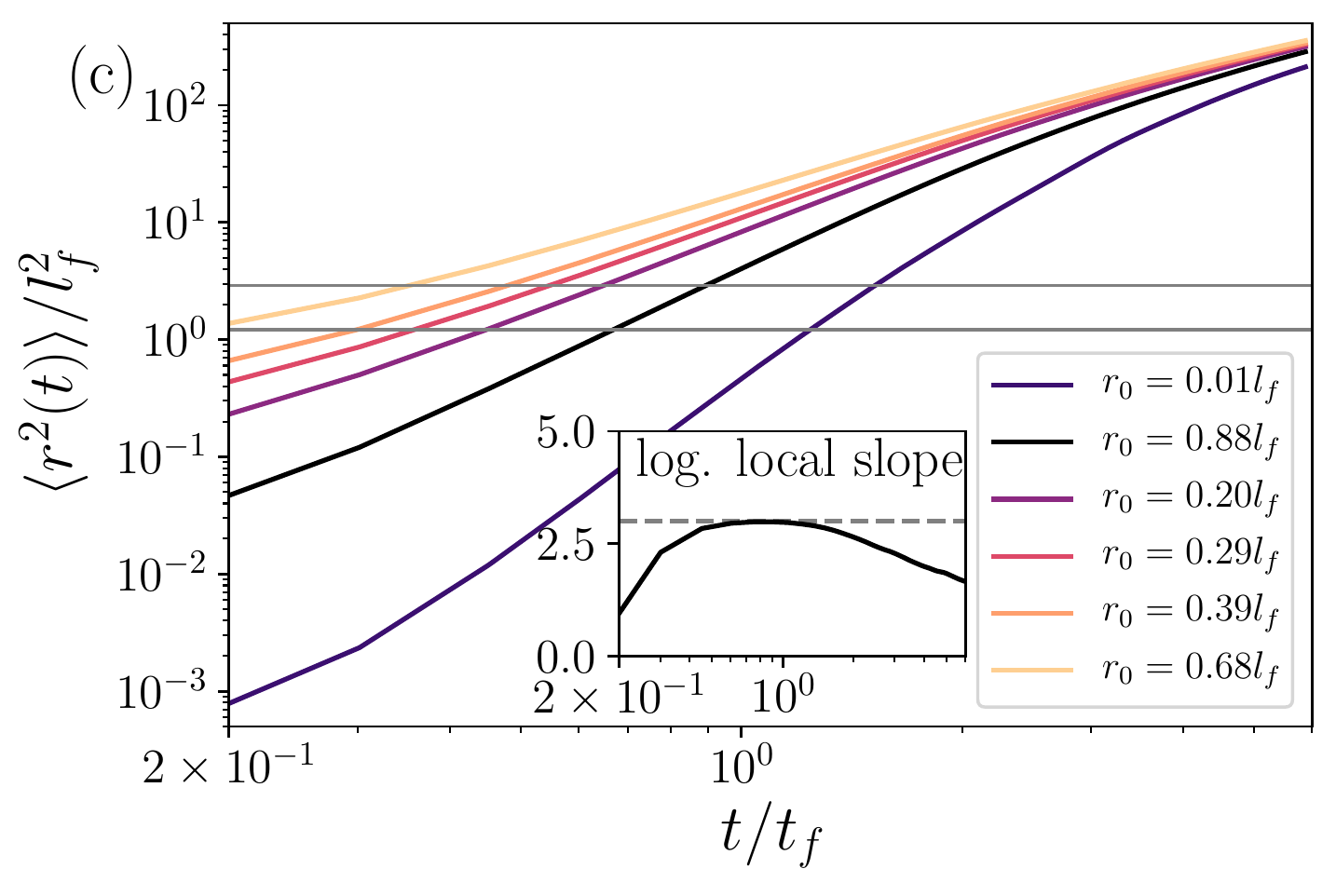}
	\includegraphics[keepaspectratio,scale=0.55]{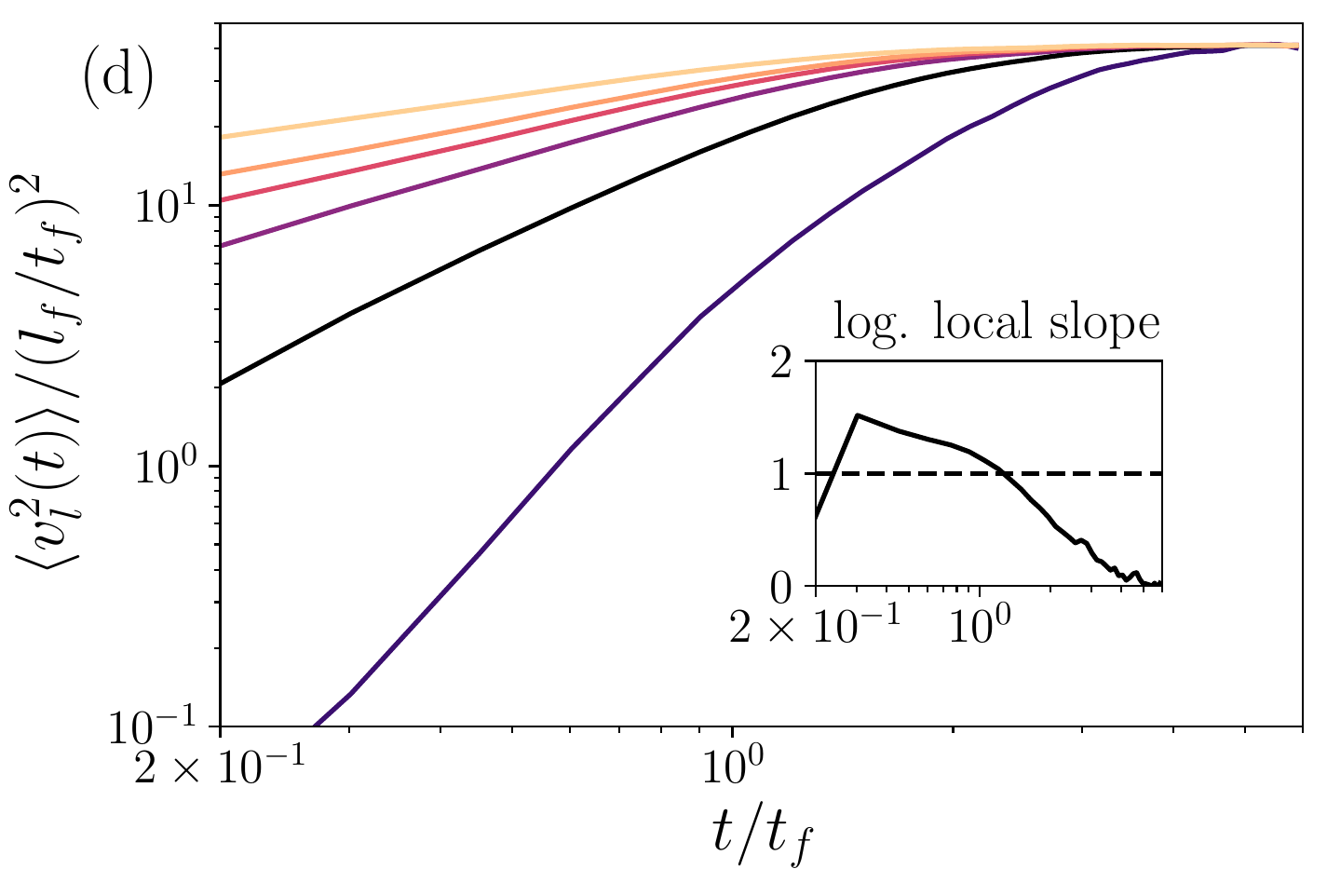}
	} 
	\caption{(a) Mean squared separation 
		for the hyperviscous case with $Re_\alpha = 40$ as a function of time
		for various initial separations.
		Here, $t_f = (l_f^2/\varepsilon)^{1/3}$ denotes the forcing time scale.
		The ranges between the two horizontal lines 
		correspond to the inertial range determined 
		by the region of the mean exit time 
		that is proportional to $r^{2/3}$, which is discussed later.
		Inset: the logarithmic local slope
		for the special initial separation, 
		$r_0 = 0.088 l_f \sim \rp$,
		where $\varepsilon$ denotes the mean energy flux 
		in the inertial range.
		The gray dashed line corresponds to 
		the Richardson scaling exponent, $3$.
		(b) mean squared relative velocity 
		for various initial separations.
		Inset: the logarithmic local slope
		for the special initial separation, $r_0 = 0.088 l_f$.
		The gray dashed line corresponds to the Kolmogorov scaling
		exponent, $1$.
		(c) Same as (a) albeit for the normal viscous case.
         (d) Same as (b) albeit for the normal viscous case. 
	}
	\label{fig:rv_2}
\end{figure}

At the Reynolds numbers currently available
in experiments and numerical simulations,  
the time evolution of $\langle r^2(t) \rangle$ depends on the initial separation.
Hence, it is not possible to conclude whether it obeys 
the Richardson--Obukhov prediction 
$\langle r^2(t) \rangle \propto \varepsilon t^3$ (
for e.g., \cite{Biferale2005,Bitane2013} for the 3D case and \cite{Jullien1999,Boffetta2002}
for the 2D inverse energy-cascade case).
The same is applicable to 
the second-order moment of relative velocity, $\langle v^2_l(t) \rangle$
for which the K41 dimensional analysis yields $\langle v^2_l(t) \rangle \propto \varepsilon t$ (for e.g., \cite{Yeung2004} for the 3D case).
In the 2D simulation at moderate Reynolds numbers, 
the initial-separation dependence is clearly
confirmed for both
$\langle r^2(t) \rangle$ shown in Fig.~\ref{fig:rv_2}(a) and (c)
and $\langle v^2_l(t) \rangle$
shown in Fig.~\ref{fig:rv_2}(b) and (d) where $t_f = (l_f^2/\varepsilon)^{1/3}$ 
denotes the forcing time scale.
We examine the results by varying the initial separations
below the forcing scale, namely $r_0 < l_f$. 
Thus, the initial separations are in 
the scales lower than the inertial range. 
If we set the initial separation in the inertial range, 
the graphs of $\langle r^2(t) \rangle$ and $\langle v^2_l(t) \rangle$ are located 
(they are not shown) above the curves plotted in Fig.~\ref{fig:rv_2}. 
Thus, we normalize all quantities by $l_f$ and $t_f$
unless there is some particular reason.
This is because $l_f$ and $t_f$ approximately define
the lowest length and time scale of the inertial range, respectively.

The data with the initial-separation dependence indicates that it is possible to select a special value corresponding to $r_0$ for which $\langle r^2(t) \rangle$ becomes
consistent with
the Richardson--Obukhov law $\langle r^2(t) \rangle = g \varepsilon t^3$. 
Further, we include the Richardson constant, $g$, which is 
non-dimensional and possibly universal.
We show the squared separation of the special case 
in the inset of Fig.~\ref{fig:rv_2}(a) as a logarithmic local slope.
However, it should be noted that (even for the special case) 
agreement of the squared velocity with the K41 prediction, 
$\langle v_l^2(t) \rangle \propto \varepsilon t$ is not as good
as that of the squared separation. 
This is observed in the inset of Fig.~\ref{fig:rv_2}(b).

Given the apparent failures of the K41,
in this study, we still argue that a certain bulk of the particle pairs starting
from each initial separation $r_0$ shown in Fig.~\ref{fig:rv_2} obey
the Richardson--Obukhov law of the squared separation even at the moderate 
Reynolds numbers.
Thus, we perform conditional sampling of particle pairs. 
The qualitative condition is that we remove particle pairs that prevent from 
separating too fast. 
In the following section, we demonstrate that this type of a conditional average
$\langle r^2(t) \rangle_c$ becomes independent of the initial separation
and that $\langle r^2(t) \rangle_c$ is the same as the unconditioned
$\langle r^2(t) \rangle$ commencing from the special initial separation
(see Fig.~\ref{fig:conditional_large_r}).
Hence, the conditional sampling recovers
the Richardson-Obukhov law, $\langle r^2(t) \rangle_c = g \varepsilon t^3$, 
including the Richardson constant and flux.
Thus, we term the special initial separation as the proper initial 
separation, which we denote as $\rp$

Evidently, our conditional sampling is contrived. 
It has several tuning parameters as we will specify them.
We determine their values empirically by ensuring that 
$\langle r^2(t) \rangle_c \propto t^3$ holds.
In order to demonstrate the extent to which it is contrived, we examine the manner in which 
conditional statistics change by varying tuning parameters.
Furthermore, we demonstrate that the number of removed pairs decreases when 
the Reynolds number increases.
The details of the conditional sampling are given in the next subsection.

\subsection{
Conditional sampling via mean exit time
\label{sec:sampling}
}
Figure~\ref{fig:rv_2}(a) plots nine cases of the different initial separations.
To develop the conditional sampling, we first focus on initial 
separations that satisfy $r_0 > \rp$, where $\rp$ denotes the proper initial separation.
Thus, we consider three cases from above in In Fig.~\ref{fig:rv_2}(a).
Our estimate of the proper initial separation is empirical:
we examine the compensated plot of the unconditional moment $\langle r^2(t) \rangle$ 
as shown in Fig.~\ref{fig:rv_2}(a) by changing $r_0$. Subsequently, we select $r_0$ for which 
the compensated plot exhibits the widest plateau. 
We evaluate the proper initial separation in this manner as 
$\rp = 0.088 l_f, 0.101 l_f$, and $0.121 l_f$ for $Re_\alpha = 40, 80$, 
and $160$, respectively.

With respect to the initial separations $r_0 > \rp$, 
the graphs of the mean squared separation $\langle r^2(t) \rangle$ 
are situated above the graph starting from $\rp$ as shown in Fig.~\ref{fig:rv_2}.
This indicates that 
it is necessary to remove particle pairs that separate too fast to recover the $t^3$ scaling.
Now, we pose two questions on the conditional sampling:
(A) Is it possible to instantaneously determine whether it is excessively fast, i.e., exhibiting excessively large $r(t)$?; 
(B) How to draw a line between excessively fast pairs and not excessively fast pairs, i.e., the threshold level between the two sets?
We handle both the questions with exit-time statistics that are
proved as effective tools in the study of relative diffusion.

\begin{figure}
	\centerline{\includegraphics[keepaspectratio,scale=0.55]{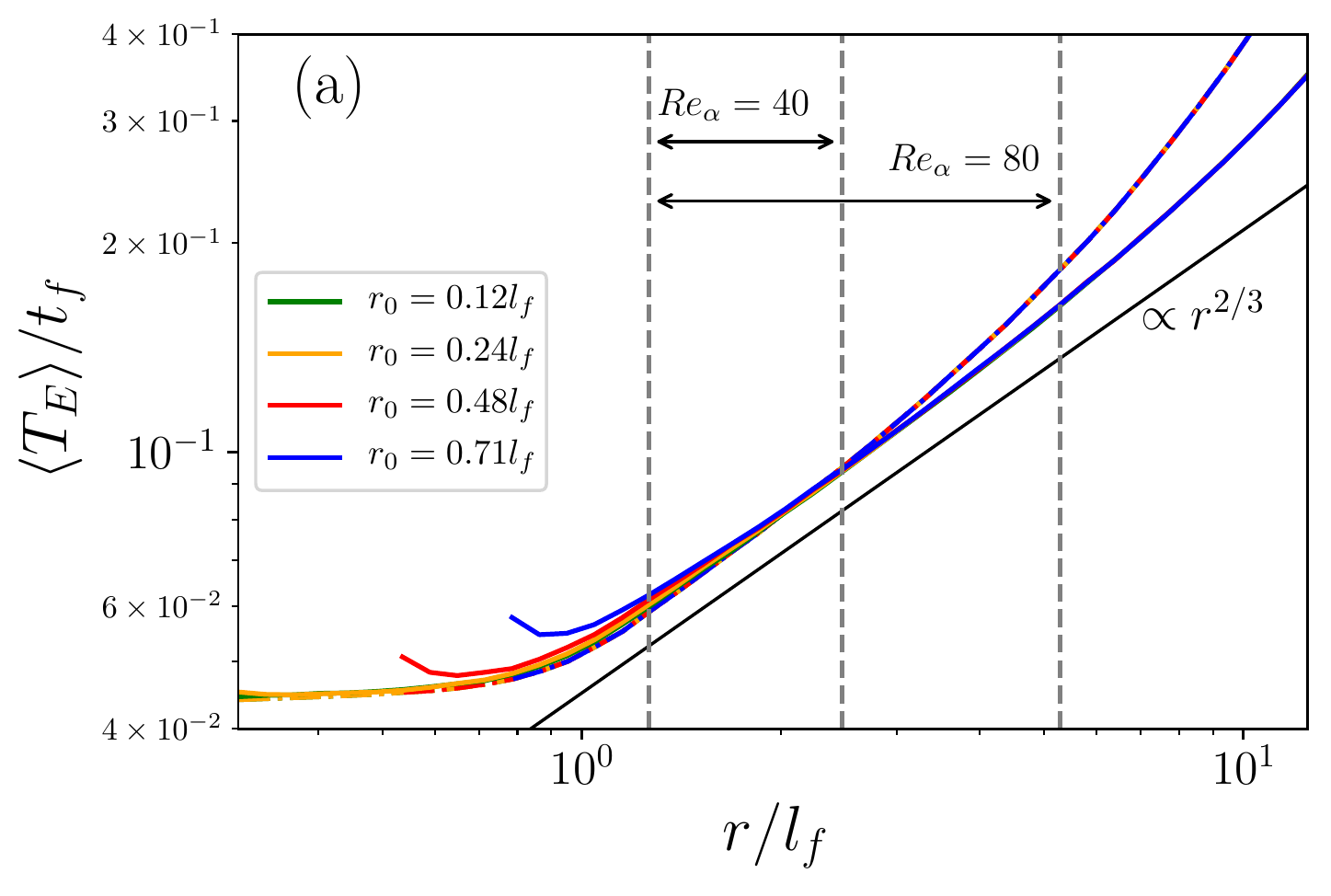}
	\includegraphics[keepaspectratio,scale=0.55]{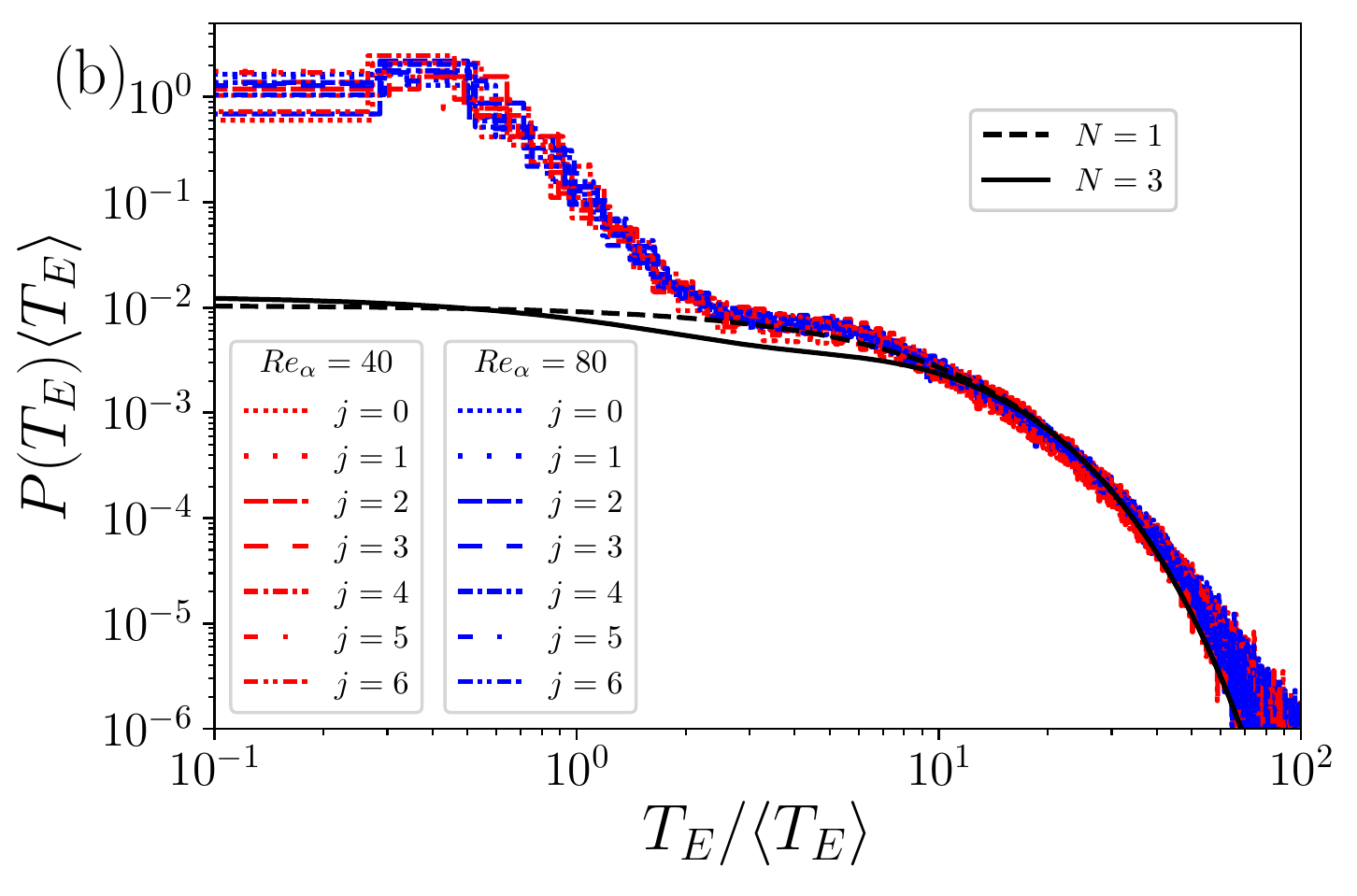}}
	\caption{(a) Mean exit time  
	for the hyperviscous cases with $Re_\alpha = 40$ (dash-dotted line)  
	and $Re_\alpha = 80$ (solid line).
	Black solid line denotes $r^{2/3}$ power law.
	The ranges between the two gray dashed lines denote 
	the inertial range estimated with the exit time 
	at $Re_\alpha = 40$ and $Re_\alpha = 80$.
	(b) The PDFs of exit time normalized by the mean at 
	$Re_\alpha = 40$ (red) and $80$ (blue) for $j=0$ to $6$.
	Black dashed line and solid line denote longtime asymptotic forms
	for the PDF of exit times, i.e.,
	$P(T_E)\langle T_E\rangle \sim C\sum_{i=1}^N j_{2,i}J_2^\prime(j_{2,i})\exp[-\frac{j_{2,i}^2}{12}\frac{\rho^{2/3}-1}{\rho^{2/3}}  \frac{t}{\langle T_E \rangle}]$
	\cite{Boffetta2002}, at $N=1$ and $N=3$, respectively. 
	Here, $C$ denotes the normalized factor, $J_2$ denotes the Bessel function, and 
	$j_{2,i}$ denotes the $i$-th zero point of $J_2$ and $\rho = 1.1$.
	}
	\label{fig:exittime}
\end{figure}

The exit time concerns the first passage time. 
The first passage time of the separation $r(t)$ for a given value $R$
is defined by the first instance when the separation $r(t)$ becomes equal to $R$.
(for the first passage time of a general stochastic process, see, e.g., \cite{Gardiner2009}).
We express the first passage time as $T_F(R)$. To define the exit time, it is necessary to 
set the domains. We denotes the domain as a series $R_0, R_1, R_2, \ldots$.
The exit time of the $j$-th zone $R_{j-1} \le r(t) < R_{j}$ is then defined 
as $T^{(j)}_E  = T_F(R_{j}) - T_F(R_{j-1})$ where $R_j = r_s \rho^{j}$ 
with parameters $r_s$ and  $\rho > 1$ for $j = 1, 2, \ldots$.
In the relative diffusion problem, exit-time statistics are introduced 
to solve the finite-size problems \cite{Artale1997,Boffetta1999,Boffetta2002,Rivera2005}.
By selecting thresholds $R_j$s in the inertial range, 
it is possible to exclusively extract information of the inertial range.
It is known that exit-time statistics 
are insensitive to the Reynolds number(for e.g.,\cite{Ogasawara2006c}). 
It is also known that the mean exit time is consistent with the K41 
prediction, $\langle T^{(j)}_E \rangle \propto R_j^{2/3}$ 
when $R_j$ is in the inertial range. Furthermore, the scaling 
behavior holds independent of the initial separations 
\cite{Boffetta2002,Biferale2005}. 
In our simulation with $r_s = 1.3 l_f$ and $\rho = 1.1$, 
the K41 scaling of the mean exit time is observed for $1 \le j \le 7$  for the hyperviscous $Re_\alpha = 40$ case
and $1 \le j \le 14$ for the hyperviscous $Re_\alpha = 80$ case 
as shown in Fig.~\ref{fig:exittime}. 
Typical value of $\rho$ as used in the previous studies 
corresponded to $1.1$ or $1.2$ \cite{Artale1997,Boffetta1999,Boffetta2002,Rivera2005,Ogasawara2006c} 
and the properties of exit-time statistics as mentioned above do not change 
in the range of $\rho$.

Now, we describe how we address the two questions of conditional sampling with the exit time.
For the first question (A), we assume that it is not possible to instantaneously determine excessively-fast
pairs. 
This is performed over certain consecutive zones in the inertial range. 
We express the number of the zones by $N_Q$
(at the end of Sec.~\ref{sec:sampling}. 
We change the parameter $N_Q$ and discuss question (A).).

With respect to the second question (B), evidently small exit time $T_E^{(j)}$ corresponds
to pairs separating fast. Hence, to remove the excessively-fast pairs, we set an upper 
threshold, $\tau$, in terms of the exit time over the zones $j = 1, 2, \ldots, N_Q$. 
Hence, if the exit time of the particle pair satisfies 
\begin{equation}
	\frac{T_E^{(j)}}{\langle T_E^{(j)} \rangle} \le \tau
	\label{eq:ex_conditions}
\end{equation}
in \textit{all} of the zones, $j = 1, 2, \ldots, N_Q$,
then this type of a pair is removed
from the Lagrangian average.
It should be noted that the threshold $\tau$ is independent of $j$.
The condition implies that the removed pairs spend a short time when compared to the average
in any of the $N_Q$ zones. Thus, the removed pair separate too fast 
in all of the monitored zones.
Conversely, the remaining pairs in the conditional sampling
generally spend a sufficiently long time 
such that $T_E^{(j)} / \langle T_E^{(j)} \rangle > \tau$.
However, they can become excessively-fast in several (but not all) of the $N_Q$ zones.
It should be noted that the conditional sampling includes two parameters: $N_Q$ and $\tau$.

Our physical picture of the conditional sampling is as follows:
The pair separation $r(t)$
is given by the time integral of the relative velocity from time $0$ to $t$.
The accumulating nature of $r(t)$ suggests that it is necessary to consider the history of a pair in conditional sampling. 
We consider it in terms of the $N_Q$ zones starting from the lowest 
scale of the inertial range. 
An actual value of $N_Q$ will be determined empirically.
With respect to $\tau$, it is noted that the right part of the PDF of the exit time is 
given by the Richardson PDF of the separation, 
$P(r, t) \propto \varepsilon^{-1} t^{-3} \exp[- ({\rm const.}) \varepsilon^{-1/3}t^{-1} r^{2/3}]$, which
denotes the self-similar solution to the Richardson's diffusion (Fokker-Plank) equation \cite{Boffetta2002}.
The correspondence of the PDFs shown in Fig.~\ref{fig:exittime}(b) indicates that the pairs in the left part
in the exit-time PDF should be removed in the conditional sampling,
thereby leading to the criterion, Eq.~(\ref{eq:ex_conditions}). 
This picture is only qualitative in nature. 

\begin{figure}
	\centerline{
	\includegraphics[keepaspectratio,scale=0.55]{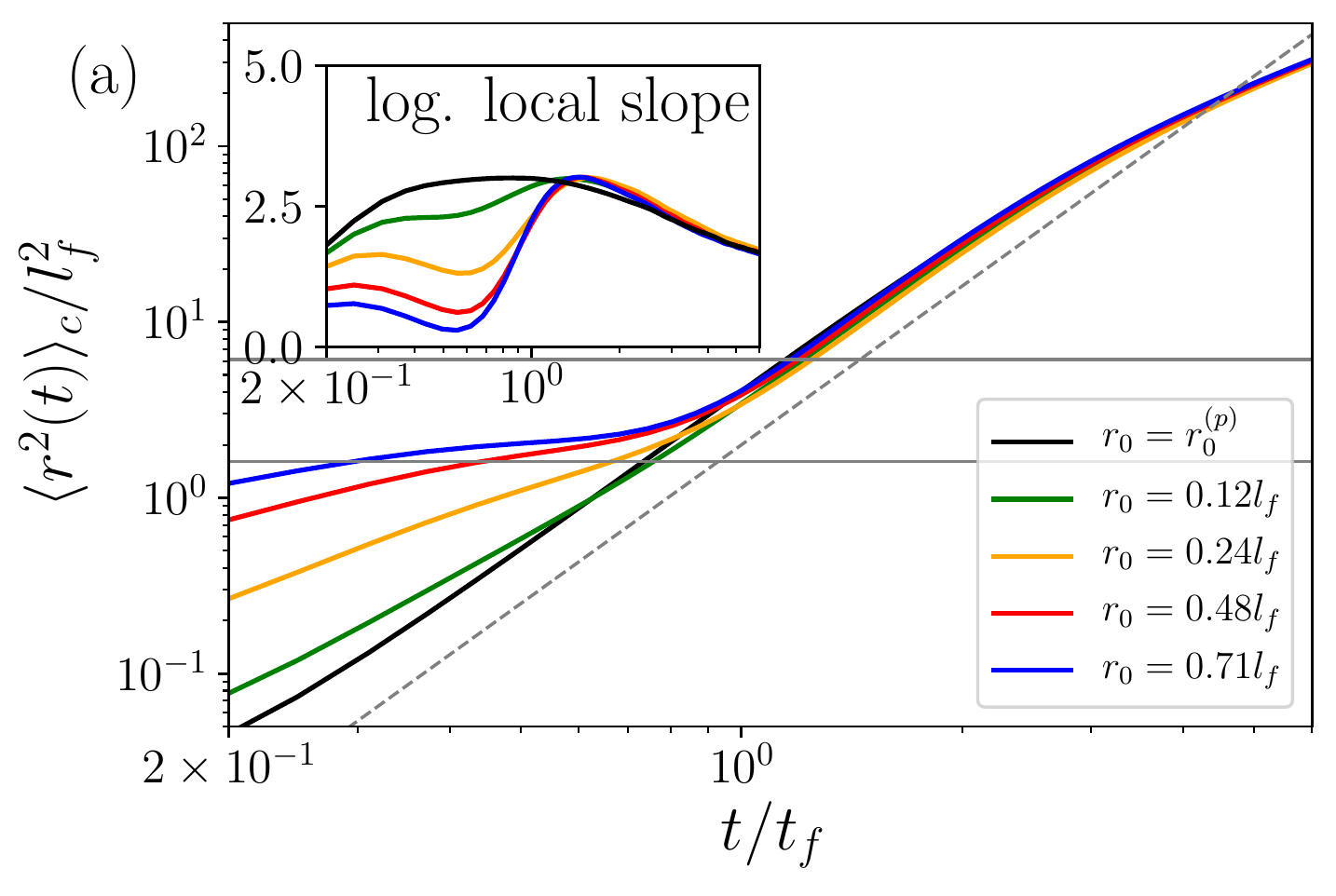}
	\includegraphics[keepaspectratio,scale=0.55]{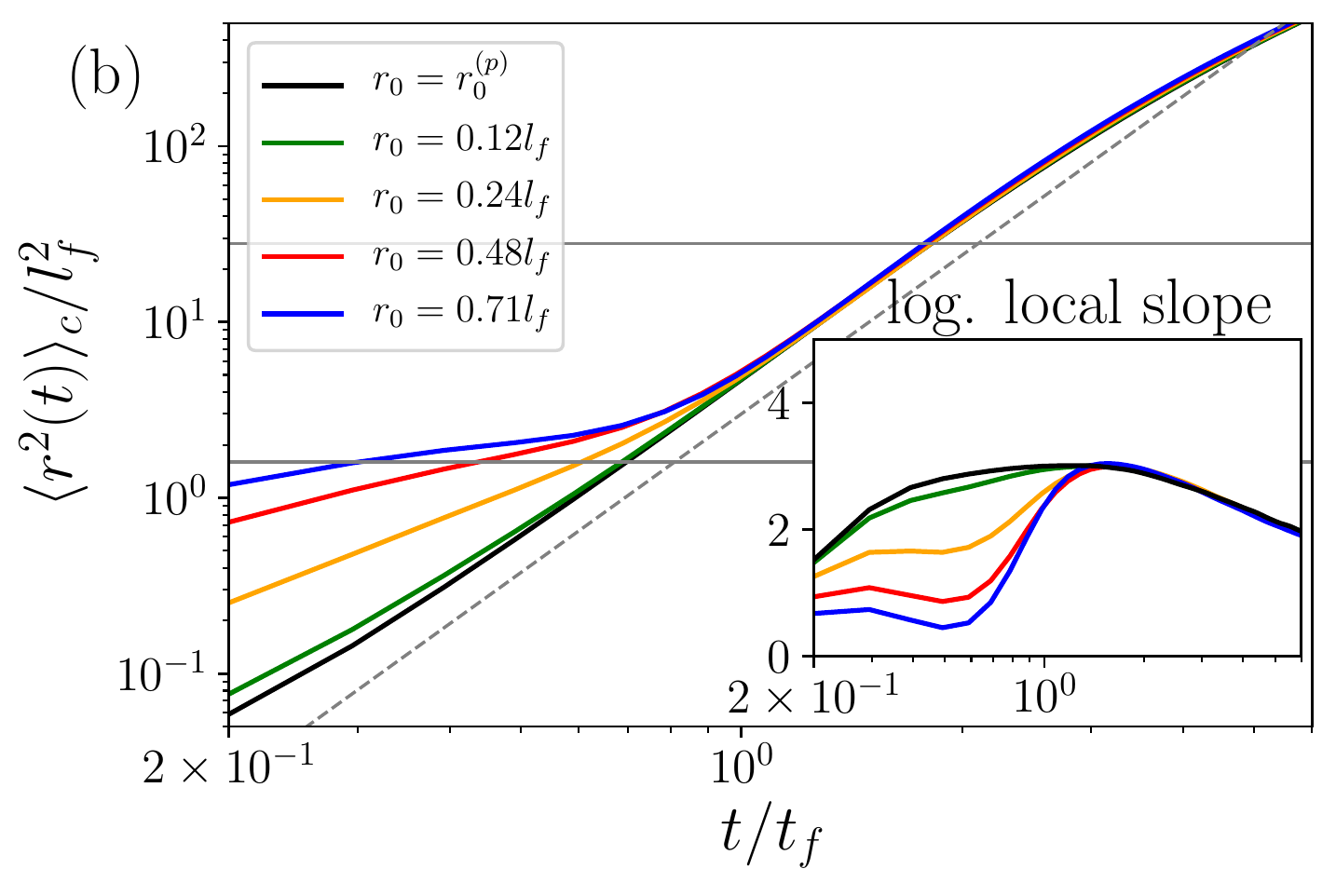}}
 \centerline{
	\includegraphics[keepaspectratio,scale=0.55]{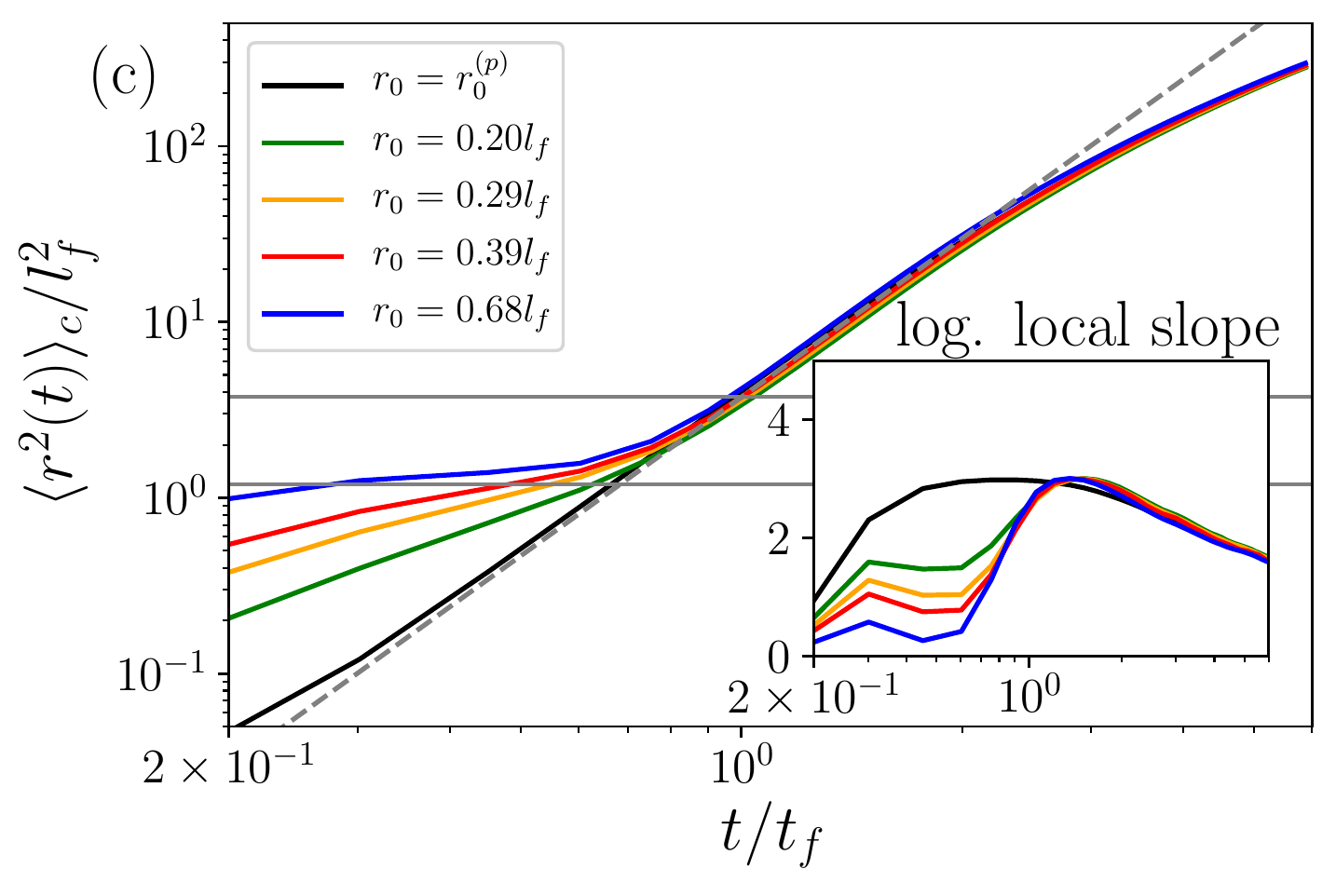}} 
 \caption{Conditionally sampled second-order moments of 
	 relative separations. The initial separations exceed the proper initial separation $\rp$.
	 (a) Conditional sampling for the hyperviscous  
	 $Re_\alpha = 40$ case
	 starting from various initial separations.
	 (b) Conditional sampling for the hyperviscous 
	 $Re_\alpha = 80$ 
	 starting from  various initial separations.
 	 (c) Conditional sampling for the viscous  
	 $Re_\alpha = 39$ case
	 starting from various initial separations. 
	 Inset shows the logarithmic local slopes 
	 of the conditionally sampled 
	 $\langle r^2(t) \rangle$.
	 Black solid curve shows 
	 the second-order moment of the relative separation 
	 without any conditional sampling 
	 starting from the proper initial separation, $r_0=\rp$. 
 	 The gray dashed line denotes $t^3$ power law. 
	 The range between the two horizontal gray solid lines 
 	 denotes the inertial range estimated with the exit time.
}
	\label{fig:conditional_large_r}
\end{figure}

We then describe the determination of parameter values in practice.
In the case of $Re_\alpha = 40$, the inertial range is covered by $7$ zones. 
Hence, we set the number of the monitored zones to $N_Q = 7$. Subsequently, we tune
the threshold $\tau$'s based on the initial separation $r_0$ to recover 
the Richardson--Obukhov law by examining whether the compensated plot 
$\langle r^2(t) \rangle_c / (\varepsilon t^3)$ exhibits a plateau.
We empirically determine that the $\tau$ values correspond to 
$0.40, 1.2, 3.4$, and $4.6$ for the initial separations $r_0/l_f = 0.12, 0.24, 0.48$, 
and $0.71$, respectively.
Given the parameters, we present the results of the conditional average of the 
squared separations in Fig.~\ref{fig:conditional_large_r}(a).
With respect to the hyperviscous $Re_\alpha = 80$ case, the inertial range is covered by $14$ zones.
However, we demonstrate the result with the same $N_Q = 7$ as that in the lower Reynolds number 
case to enable a better comparison in Fig.~\ref{fig:conditional_large_r}(b). 
The thresholds are observed as $\tau = 0.25, 0.50, 1.4$ and $3.0$ for the same set of 
the initial separations $r_0/l_f = 0.12, 0.24, 0.48$ and $0.71$,
respectively. In the normal-viscous case, $N_Q = 7$ and the thresholds correspond to
$\tau = 0.45, 1.05, 2.4$, and $4.4$ for the initial separations $r_0/l_f = 0.20, 0.29, 0.39$ and $0.68$, respectively.

As shown in Fig.~\ref{fig:conditional_large_r}, 
the conditioned curves $\langle r^2(t) \rangle_c$ collapse
in the inertial range and beyond the unconditioned curve 
for proper initial separation.
It should be noted that the width of the collapsed region increases as we increase $Re_\alpha$.
When we compare $\tau$ between the two Reynolds number cases for the same normalized 
initial separation $r_0/l_f$, it approximately decreases by a factor of $1/2$.
The fraction of the remaining pairs in the conditional sampling 
corresponds to $41$ \% for $Re_\alpha = 40$ and
$65$ \% for $Re_\alpha = 80$.
Qualitatively, the increase in the fraction is interpreted as follows. 
We assume that we compare each pair's distance $r(t)$ at the same time $t$
for the two Reynolds numbers. Given the wider inertial range at higher $Re_\alpha$, 
pairs with larger separation $r(t)$ (i.e., pairs separating fast) 
are tolerated in the higher $Re_\alpha$ case to recover the Richardson--Obukhov law.
The increase in the fraction supports the working hypothesis that a certain bulk of particle pairs 
obey the Richardson--Obukhov law even at moderate Reynolds numbers.

We then examine changes in the results of the conditional sampling 
when we vary parameters $N_Q$ and $\tau$'s for various initial separation $r_0$.
For the purpose of simplicity, we limit ourselves to the two hyperviscous cases with $Re_\alpha = 40$ and $80$.
With respect to the reference exit-time statistics, we do not change the parameters $r_s = 1.30 l_f$ 
and $\rho = 1.1$.
At $Re_\alpha = 40$, we use $N_Q = 7$ as the number of monitored zones 
independent of the initial separation $r_0$. The $N_Q$ zones, $R_1 =  r_s\rho \le R \le R_7 = r_s \rho^7$, 
cover almost the entire inertial range. We consider the initial separations satisfying $\rp < r_0 < r_s$.
We then reduce $N_Q$ to $6, 5, 4, 3, 2$, and $1$ although we use the same set of $\tau$'s determined with $N_Q = 7$.
The results indicate that further tuning of $\tau$ for the change in $N_Q$ is not necessary.
The reduction of $N_Q$ does not alter the behavior of $\langle r^2(t) \rangle_c$ 
as shown in Fig.~\ref{fig:conditional_large_r}(a).
The same is applicable to the higher $Re_\alpha = 80$ case 
shown in Fig.~\ref{fig:conditional_large_r}(b).
In this case, we change $N_Q$ to $14, 13, \dots, 1$ although 
we use the same $\tau$ for each $N_Q$.
Therefore, the result of the conditional sampling is robust relative to 
changes in the parameters. 
An important result obtained in the examination is that $N_Q = 1$ is sufficient.
This answers question (A) on the conditional sampling, namely it is not possible to instantaneously
determine if a given pair is excessively fast 
(consequently exhibiting excessively high $r(t)$).  
However, it can be performed in terms of the exit time of the first zone 
in the inertial range.
Thus, 
it is possible to locally remove excessively-fast pairs to recover the Richardson--Obukhov law in space at the entry of the inertial range.
Evidently, this is not locally in time.
This implies that evolution of a pair in the inertial range is somewhat monotone after the entry.
As shown in the next section, this is observed as a self-similar evolution
of the relative velocity.

\begin{figure}
 	\centerline{
	\includegraphics[keepaspectratio,scale=0.55]{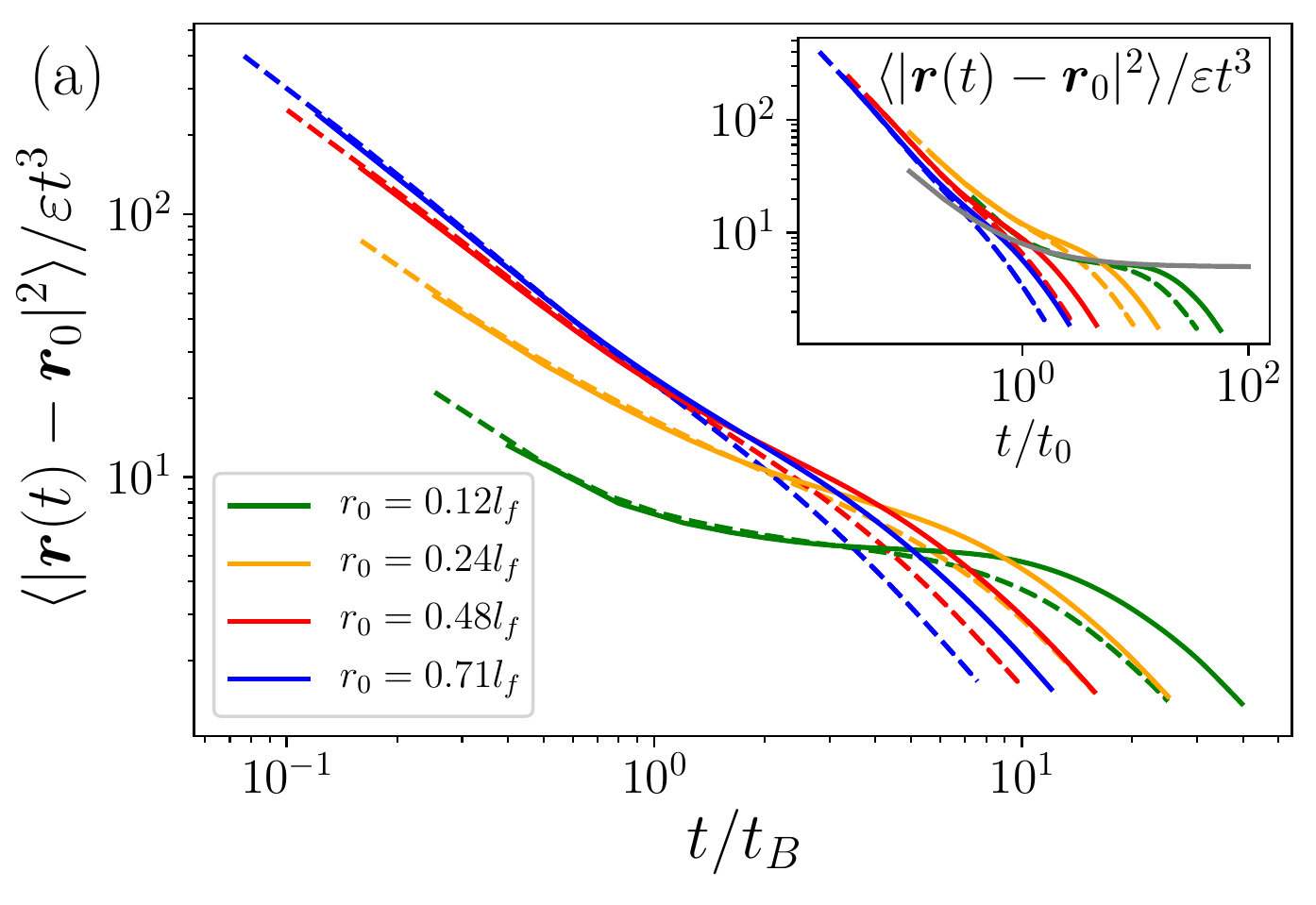}
	\includegraphics[keepaspectratio,scale=0.55]{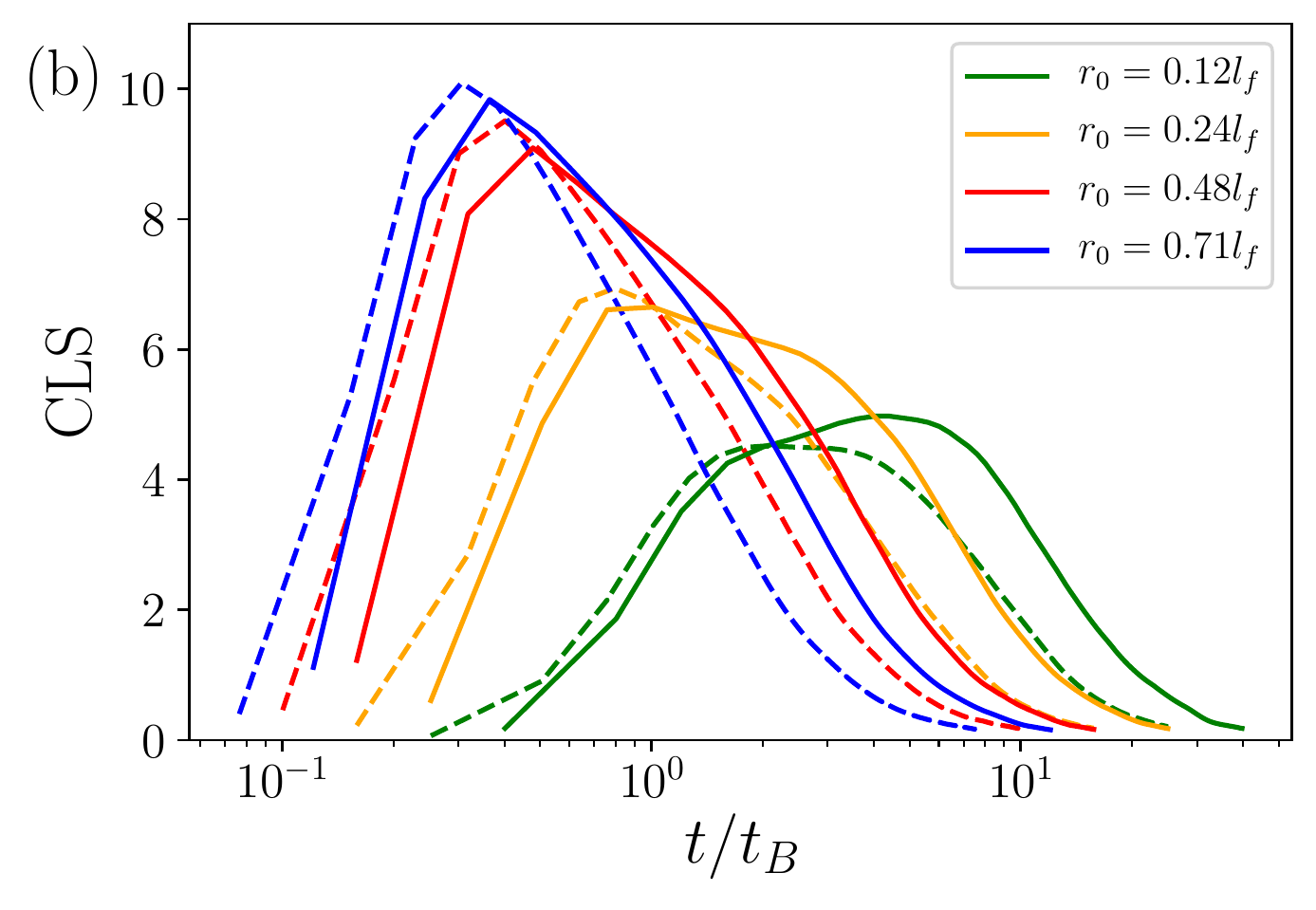}}
\caption{(a) Mean squared separations for the hyperviscous case with 
	$Re_\alpha=40$ (dashed line) and $Re_\alpha = 80$ (solid line),
	although the initial separation is 
	subtracted for various initial separations. 
	This is compensated by the scaling $\varepsilon t^3$.
	Inset shows the same plots as the main panel although
	the horizontal axis is compensated by the Bitane time scale,  $t_0 = S_2(r_0)/2\varepsilon$.
	Gray line denotes the scaling law, $\langle \bm{r}(t) - \bm{r}_0 \rangle = g\varepsilon t^3 (1+Ct_0/t)$ 
	as suggested by Bitane et al. \cite{Bitane2012}.
	Here, $C=0.6$.
	(b) Cubed-local slopes of $\langle r^2(t) \rangle$ for 
	various initial separations 
	at the hyperviscous $Re_\alpha = 40$ (dashed line) and 
	$Re_\alpha = 80$ (solid line).
	In panel (a) and (b), the time is normalized with the Batchelor time,
	$t_\mathrm{B}=r_0^{2/3}\langle \varepsilon \rangle^{-1/3}$.
	In the inset of panel (a), the time is normalized with Bitane time,
	$t_0 = S_2(r_0)/2\varepsilon$.
	}
	\label{fig:othermethod} 
\end{figure} 

Other methods are developed to remove 
the initial separation dependence.
 For the purpose of comparison, we apply two methods used 
for 3D turbulence \cite{Bitane2012, Bitane2013, D.Buaria2015}
to 2D data without utilizing conditional sampling.
A method involves subtracting the initial-separation vector 
${\bm r}_0$ from the separation vector ${\bm r}(t)$.
In Fig.~\ref{fig:othermethod} (a), 
we plot $\langle |{\bm r} - {\bm r}_0|^2 \rangle / (\varepsilon t^3)$ 
of our data for the $Re_\alpha = 80$ case with the hyperviscosity.
The Richardson--Obukhov law appears as a plateau in the region 
$t / t_B \gg 1$ or $t/t_0 \gg 1$.
Although the range of $t/t_B$ in our data is comparable to that in the 3D study \cite{D.Buaria2015},
the degree of collapse of our 2D data is worse than that of the 3D result.
The other method involves extracting the possibly subdominant $t^3$ term in $r(t)$ with a suitable exponentiation and
temporal finite difference.
In Fig.~\ref{fig:othermethod} (b), 
we plot the cubed local-slope (CLS), $ \left\{(d/dt)[\langle r^2(t) \rangle^{1/3}] \right\}^3 / \varepsilon$, 
\cite{D.Buaria2015} of our 2D data. The Richardson--Obukhov law appears as a plateau in the CLS
in $t / t_B \gg 1$. However, the degree of the collapse for the 2D result is worse.
The discrepancy between the 2D and 3D cases can be ascribed to the difference in the physics of turbulence 
in 2D and 3D.
Conversely,
the plateau is unclear irrespective of the dimensions.
This can be due to finite Reynolds number effects.
However, in order to evaluate the effects, it is necessary
to add the tuning parameter to the Richardson--Obukhov law.
A physical meaning of the tuning parameter is obscure in many cases.
Although the scaling law suggested by Bitane et al. \cite{Bitane2012}
approximately corresponds to data for finite Reynolds number
at small time, $t \lesssim t_0$,
by tuning the parameter, $C$,
it does not correspond to the data at large time, $t \gtrsim t_0$.
Hence, it is necessary to add another tuning parameter for large time.
Furthermore, the cause for the difference between $\langle r^2(t) \rangle$
and $ \langle |{\bm r}(t) - {\bm r}_0|^2 \rangle $ is not clarified.
Hence, it is insufficient to only investigate statistical moments of 
all particle pairs.
Thus, it is necessary to investigate the PDF of particle pairs.
We should consider extreme events of particle pairs 
that can affect even lower moments 
such as $\langle r^2(t) \rangle$.
It is intrinsically necessary to consider conditional statistics on
a special part of particle pairs.

So far, we restricted the conditional sampling for the 
cases of $r_0 > \rp$.
For lower initial separations, $r_0 < \rp$,
we can also recover $t^3$ scaling 
with the same conditional statistics.
However, we found that the results indicate the condition for the threshold, $\tau$, changes from
the inequality (\ref{eq:ex_conditions}) to
\begin{equation}
	\tau_1 \le \frac{T_E^{(j)}}{\langle T_E^{(j)} \rangle} \le \tau_2,
	\label{eq:ex_conditions_small}
\end{equation}
where $\tau_1 \not=0$.
For example, we empirically determine 
$(\tau_1, \tau_2) = (0.16, 7.5), (0.2, 1.0)$ and 
$(0.18, 0.50)$ for initial separations 
$r_0/l_f = 0.024 , 0.049$, and $0.073$, respectively
at $Re_\alpha = 40$.
Although $t^3$ scaling law is recovered via conditional sampling, 
the Richardson constant, 
$g = \langle r^2(t) \rangle_c/ \varepsilon t^3$,
for the conditional data is extremely sensitive to 
the initial separations (figure not shown).
The sensitivity considerably differs from the cases of $r_0 > \rp$.
This indicates that for $r_0 < \rp$ cases, we fail 
to construct conditional statistics that remove the initial separation 
dependence. We infer that in these cases the bulk of particle pairs
do not obey the Richardson--Obukhov law in the aforementioned cases. 
The initial separations are extremely small such that the pairs experience
effects from the dissipation range and the small-scale forcing is longer
than that of the cases with $r_0 > \rp$. 
Hence, two parameters are required for conditional sampling.
We do not focus on cases with $r_0 < \rp$ in the remaining part of the paper.
However, the failure implies that $\rp$ corresponds to the border line of the initial separation,
beyond which the bulk of the particle pairs becomes consistent with the Richardson--Obukhov law. 

\section{
Scaling of the relative velocity
\label{sec:results}
}

\subsection{Conditional sampling}

\begin{figure}
\centerline{
	\includegraphics[keepaspectratio,scale=0.55]{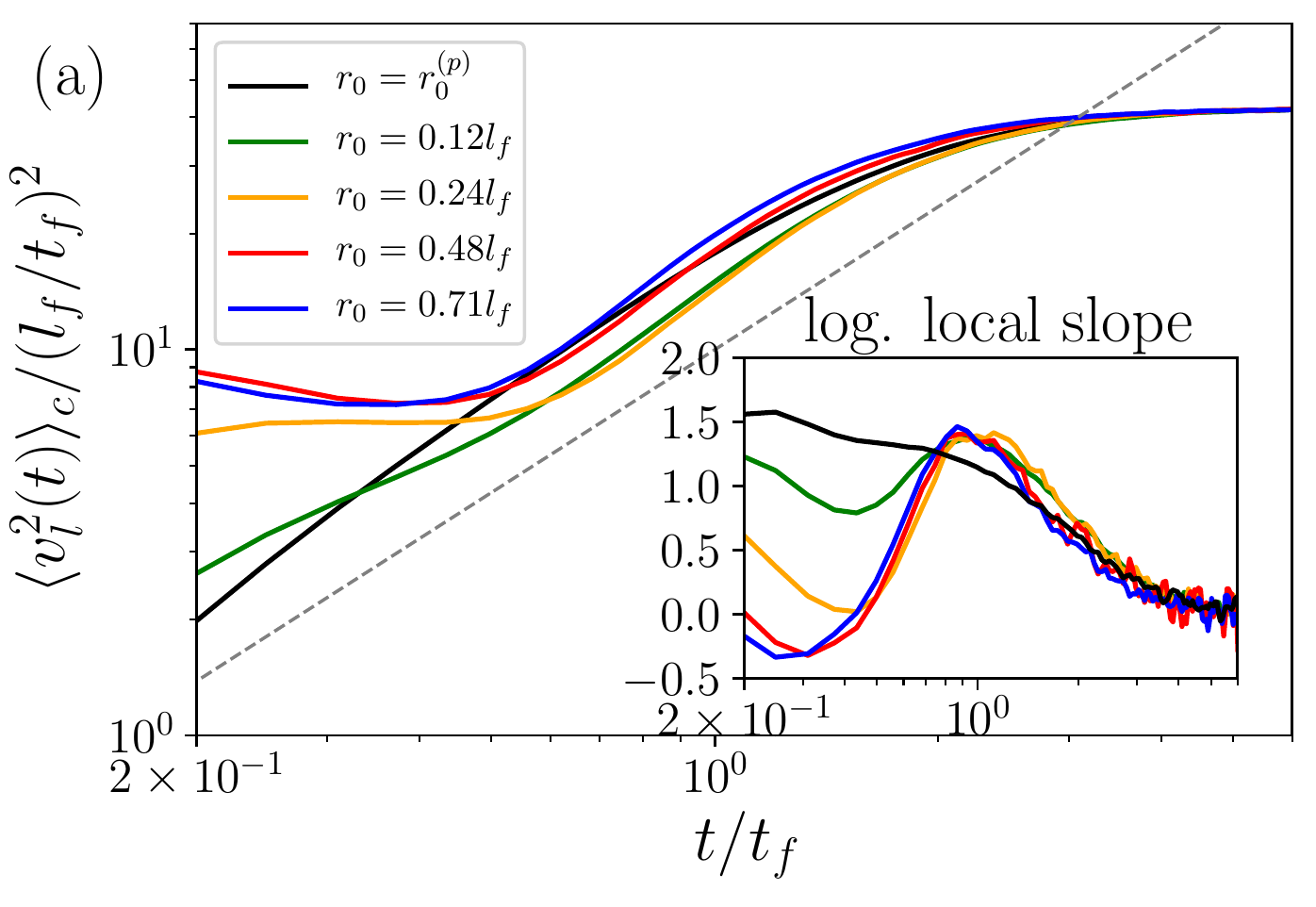}
	\includegraphics[keepaspectratio,scale=0.55]{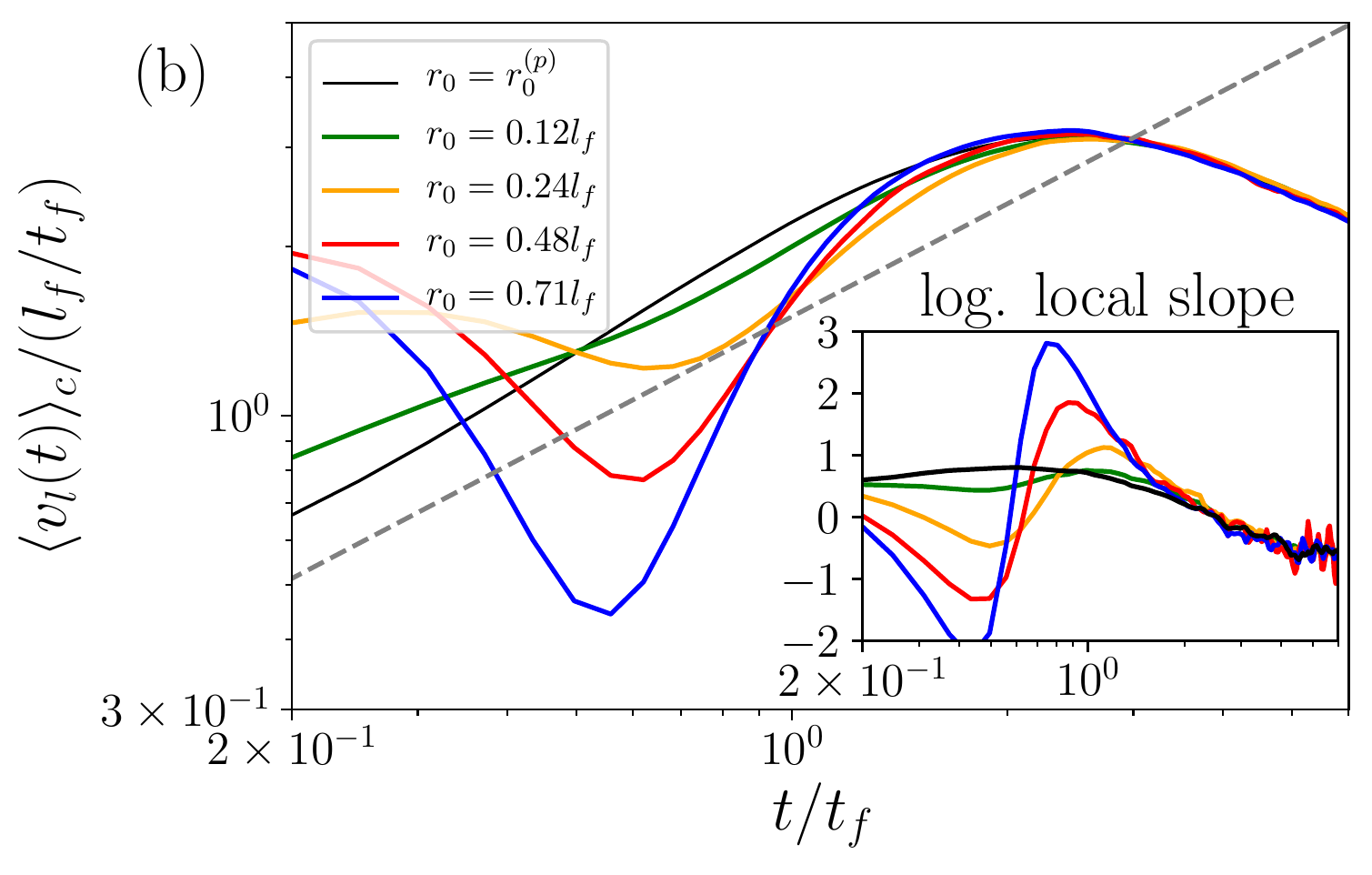} 
}
\centerline{
	\includegraphics[keepaspectratio,scale=0.55]{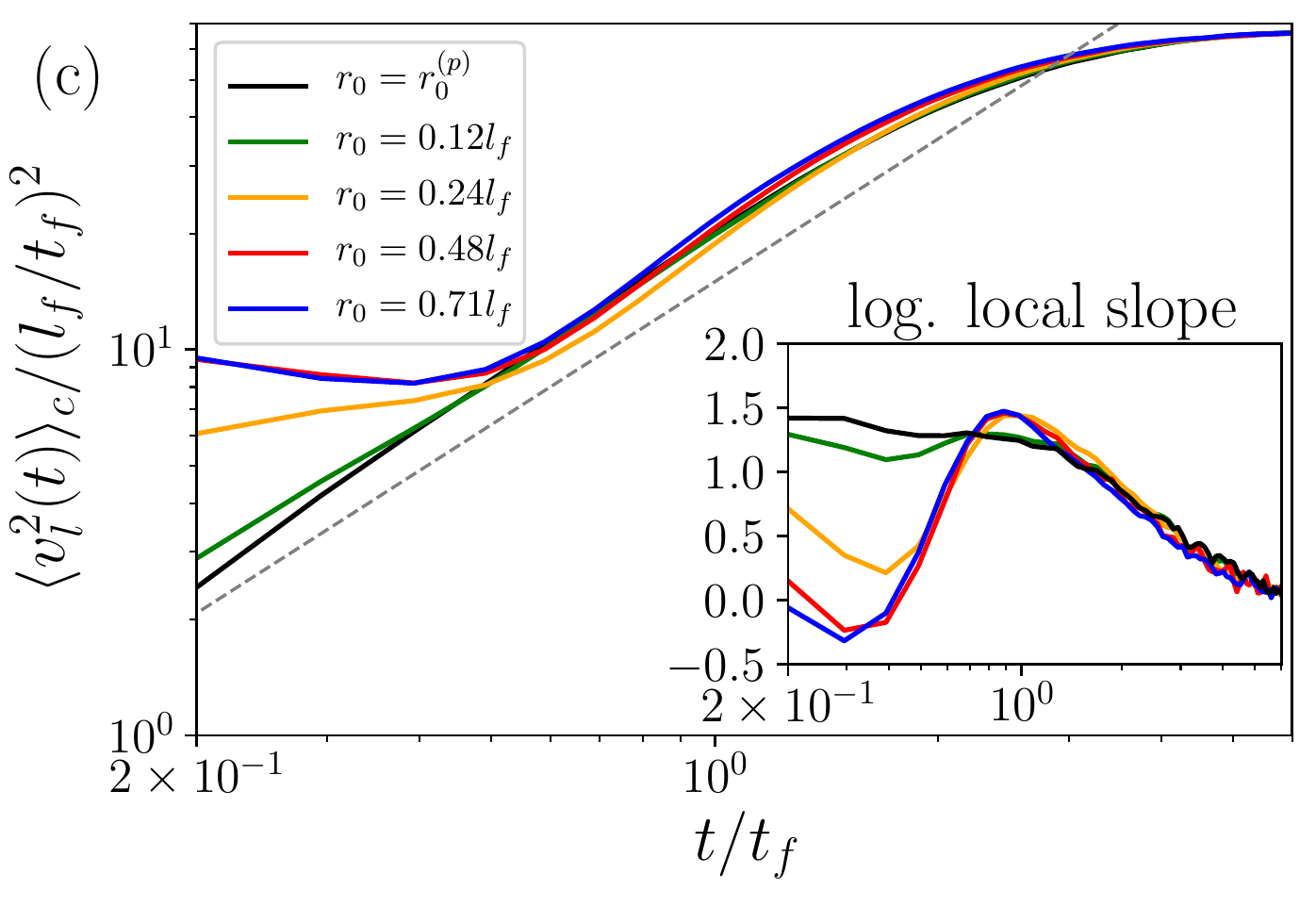}
	\includegraphics[keepaspectratio,scale=0.55]{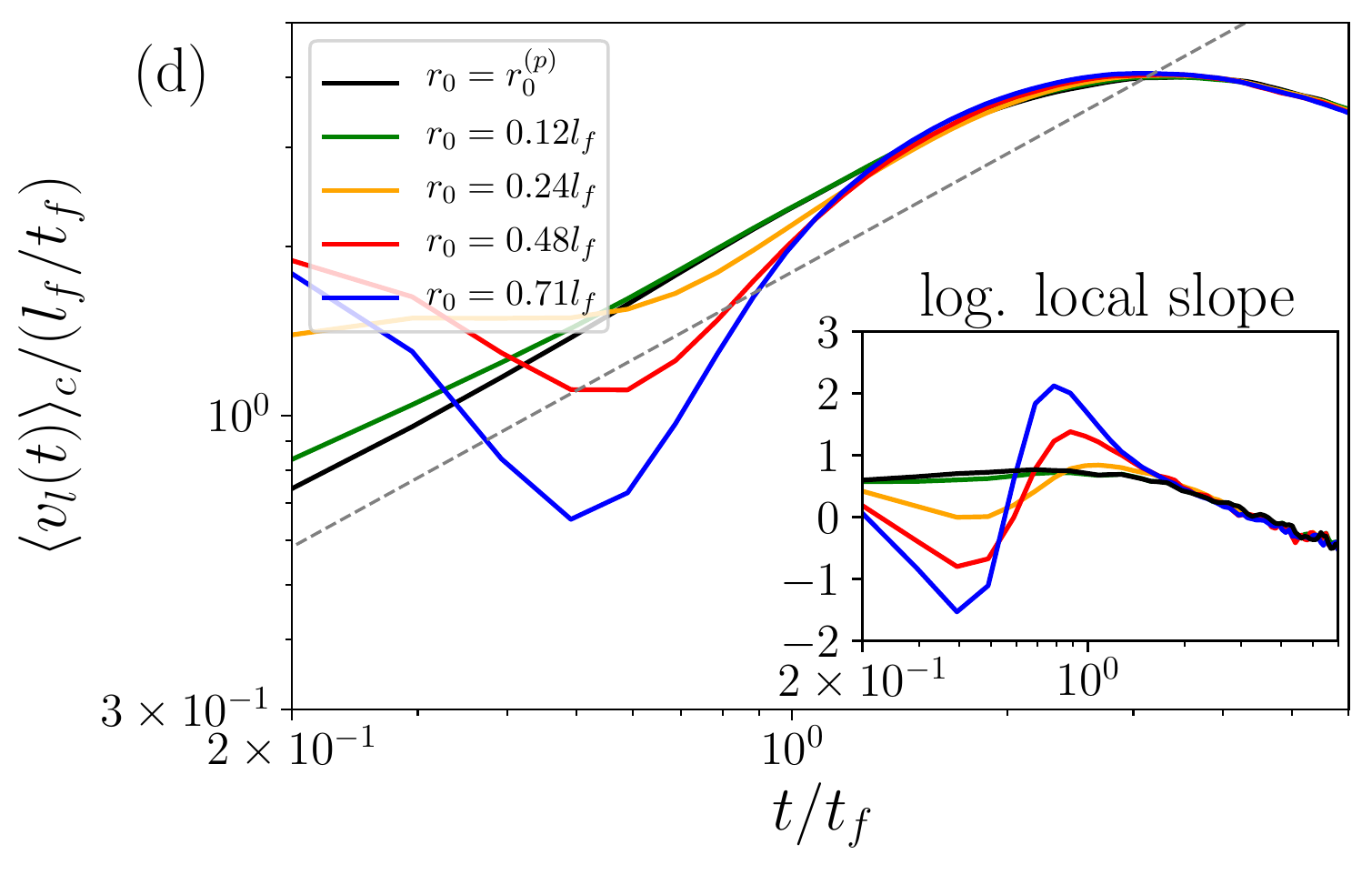} 
}
\centerline{
	\includegraphics[keepaspectratio,scale=0.55]{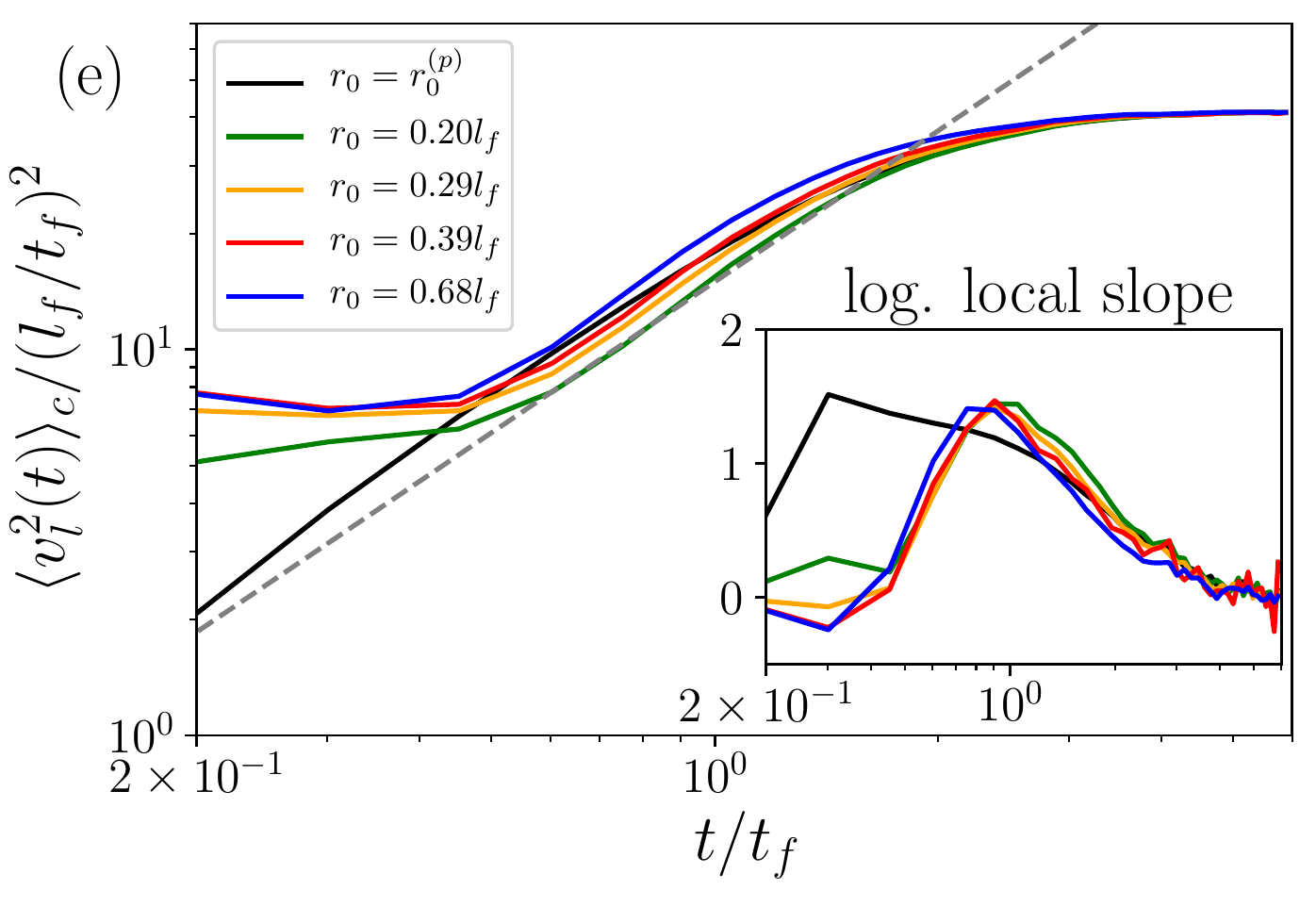}
	\includegraphics[keepaspectratio,scale=0.55]{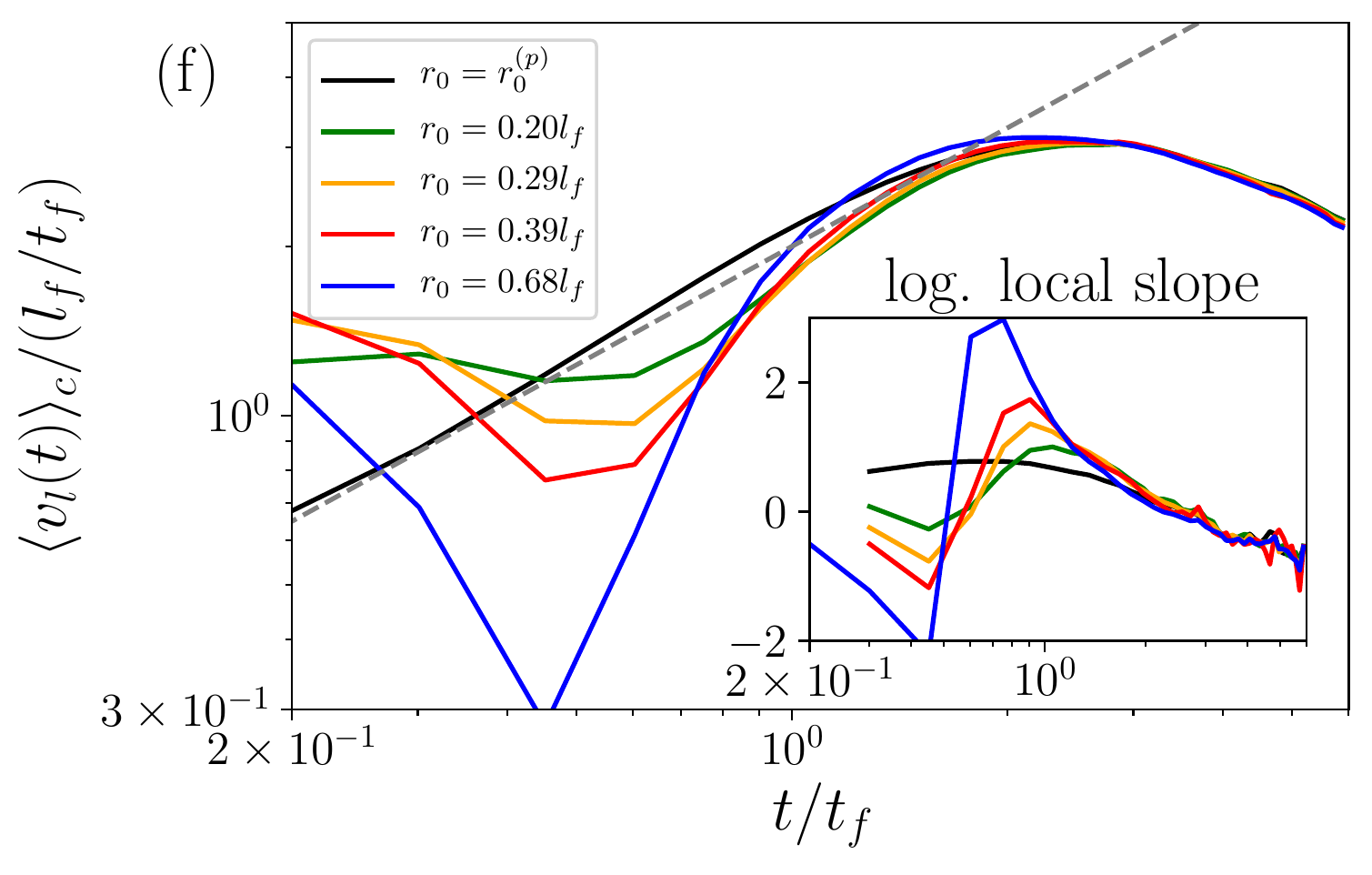}
}
 \caption{
	 Second-order (left panels) and first-order (right panels) moments
	 of the longitudinal relative velocity 
	 for conditionally sampled data
	 starting from various initial separations
	 and for the unconditioned data
	 starting from the proper initial separation, $\rp$.
	 Inset: Logarithmic local slope.
	 (a) (b) $Re_\alpha = 40$ with hyperviscosity.
	 (c) (d) $Re_\alpha = 80$ with hyperviscosity.
	 (e) (f) $Re_\alpha = 39$ with normal viscosity.
	 Dashed line denotes $t^{1.23}$ and $t^{0.7}$ scalings
	 for second-order and first-order moments, respectively.
 }
 \label{fig:moment_v}
\end{figure}

Using the conditional sampling described in the previous section,
we show conditional averages of 
the squared longitudinal relative velocity, 
$\langle v^2_l(t) \rangle_c$ in Fig.~\ref{fig:moment_v}.
Conditional velocity statistics exhibit a collapse 
similar to that of the conditional separations.
Hence, the second-order and first-order conditional moments,
$\langle v^2_l(t) \rangle_c$ and $\langle v_l(t) \rangle_c$, respectively, 
starting from various initial separations become almost identical to 
the unconditioned moments starting from the proper initial separation.
However, the degree of collapse of the velocity data 
is worse at the hyperviscous $Re_\alpha = 40$ and normal viscous $Re_\alpha = 39$ although 
it improves at $Re_\alpha = 80$ with hyperviscosity.
 The results indicate that the second-order conditional moment
$\langle v^2_l(t) \rangle_c$ and conditional average 
$\langle v_l(t) \rangle_c$ deviate from their K41 power-law 
predictions, $t^1$ and $t^{1/2}$, respectively. 
This contrasts with the conditional relative separation 
$\langle r^2_l(t) \rangle_c$ that is driven as 
consistent with the K41 prediction or the Richardson--Obukhov law.

\begin{figure}
\centerline{
	\includegraphics[keepaspectratio,scale=0.55]{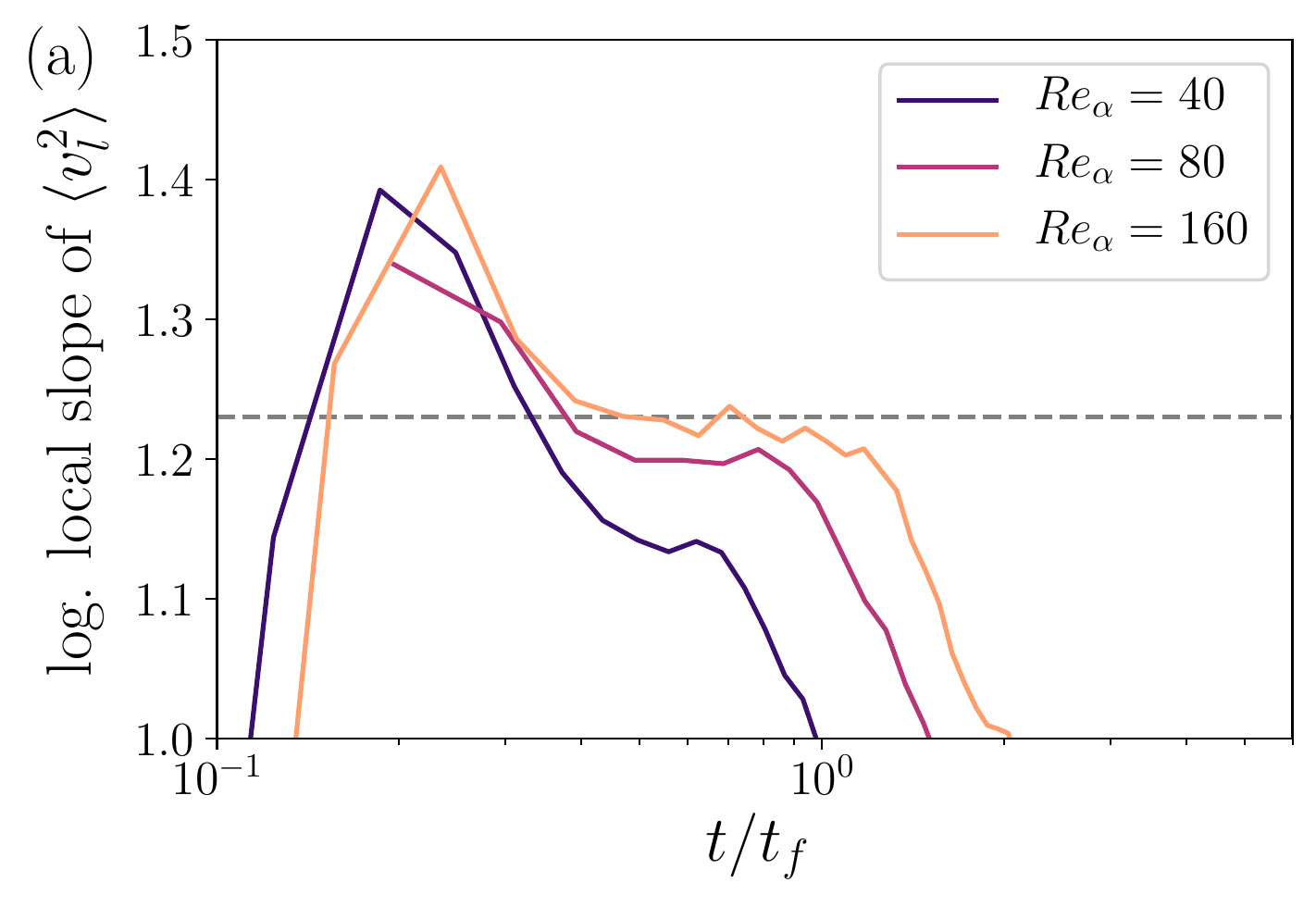}
	\includegraphics[keepaspectratio,scale=0.55]{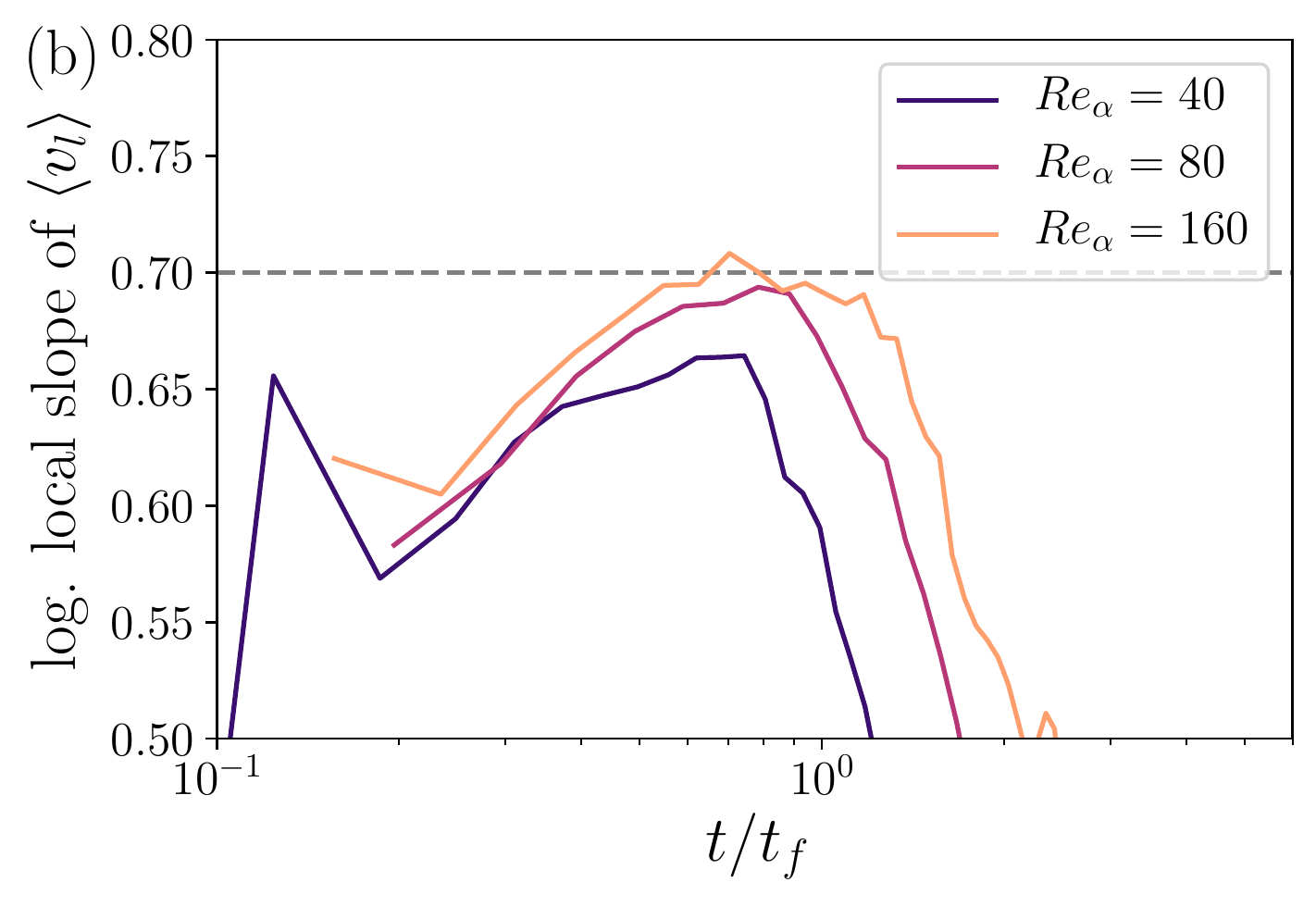}
}
 \caption{
	 $Re_\alpha$-dependence of
	 the logarithmic local slope of (a) $\langle v_l^2(t) \rangle$,
	 and (b) $\langle v_l(t) \rangle$,
	 for unconditioned data at the proper initial separation, $\rp$
	 at $Re_\alpha = 40, 80$, and $160$.
	 Gray dashed lines correspond to 
	 (a) $1.23$ and (b) $0.70$, respectively.
 }
 \label{fig:slope_v}
\end{figure}

We observe the deviation from the Kolmogorov scaling exponents and then measure the
exponents from the logarithmic local slopes of $\langle v_l^2(t) \rangle$ and $\langle v_l(t) \rangle$
shown in the insets of Fig.6. At large times, the converging behavior of the local slopes to that
of the proper initial separation is observed. 
However, a plateau is absent in the converged part.
We then assume that at higher $Re_\alpha$, the converged part
corresponds to plateau and that the level of the converged (hypothetical) 
plateau is identical to that of the proper initial separation. 
We then plot the logarithmic local slopes of 
the data starting from the proper initial separation 
with three $Re_\alpha$s in Fig.7. 
We observe that increases in $Re_\alpha$ widen the plateau and that the levels 
of the plateaus do not approach the K41 scaling exponents, 
which correspond to the bounds of vertical axis in Fig.\ref{fig:slope_v}. 
Furthermore, it should be noted that
the differences between the neighbor levels
decreases when $Re_\alpha$ increases. 
This indicates that asymptotic exponent values are present.
As shown in Fig.\ref{fig:slope_v}, given our assumptions of the converged behavior, we infer that the scaling exponents 
of the velocity statistics are
\begin{align}
	\langle v_l(t) \rangle &\propto t^{0.7}, \label{eq:v1_scaling}\\
	\langle v_l^2(t) \rangle &\propto t^{1.23}. \label{eq:v2_scaling}
\end{align}
The scaling exponents are visually determined from Fig.\ref{fig:slope_v}.
The values increase with increases in $Re_\alpha$.

\begin{figure}
\centerline{
	\includegraphics[keepaspectratio,scale=0.55]{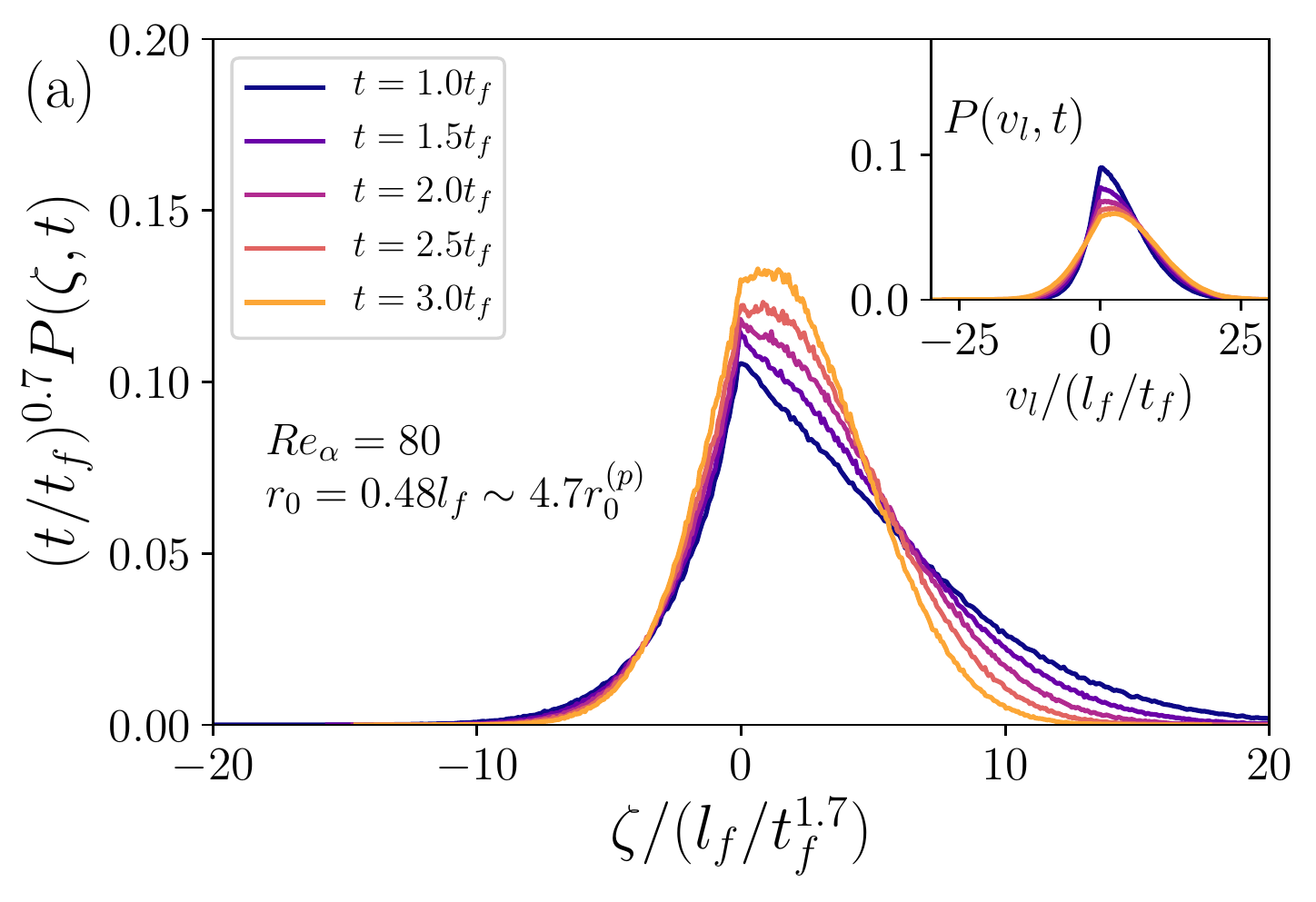}
	\includegraphics[keepaspectratio,scale=0.55]{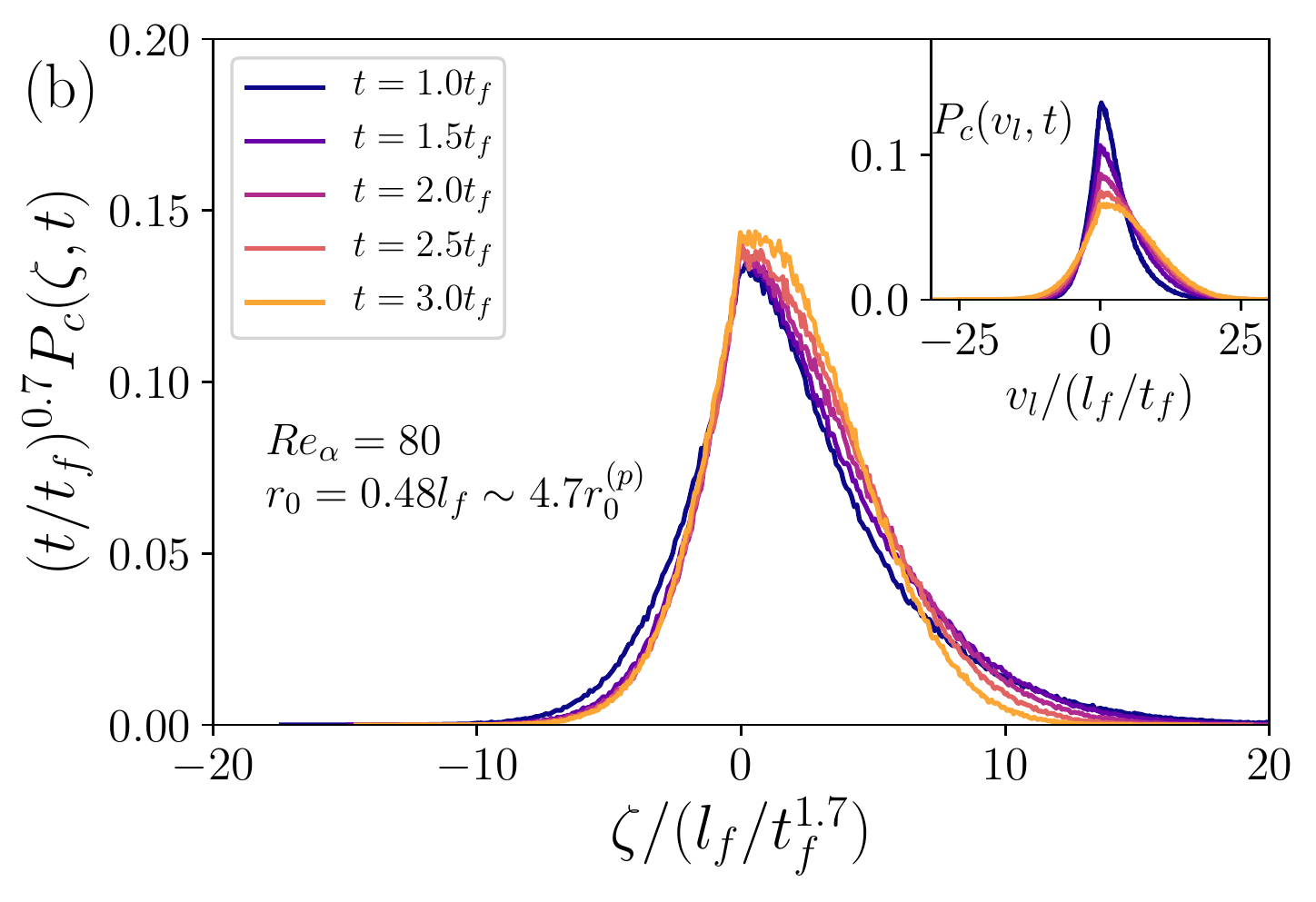}
	}
	\centering{
	\includegraphics[keepaspectratio,scale=0.55]{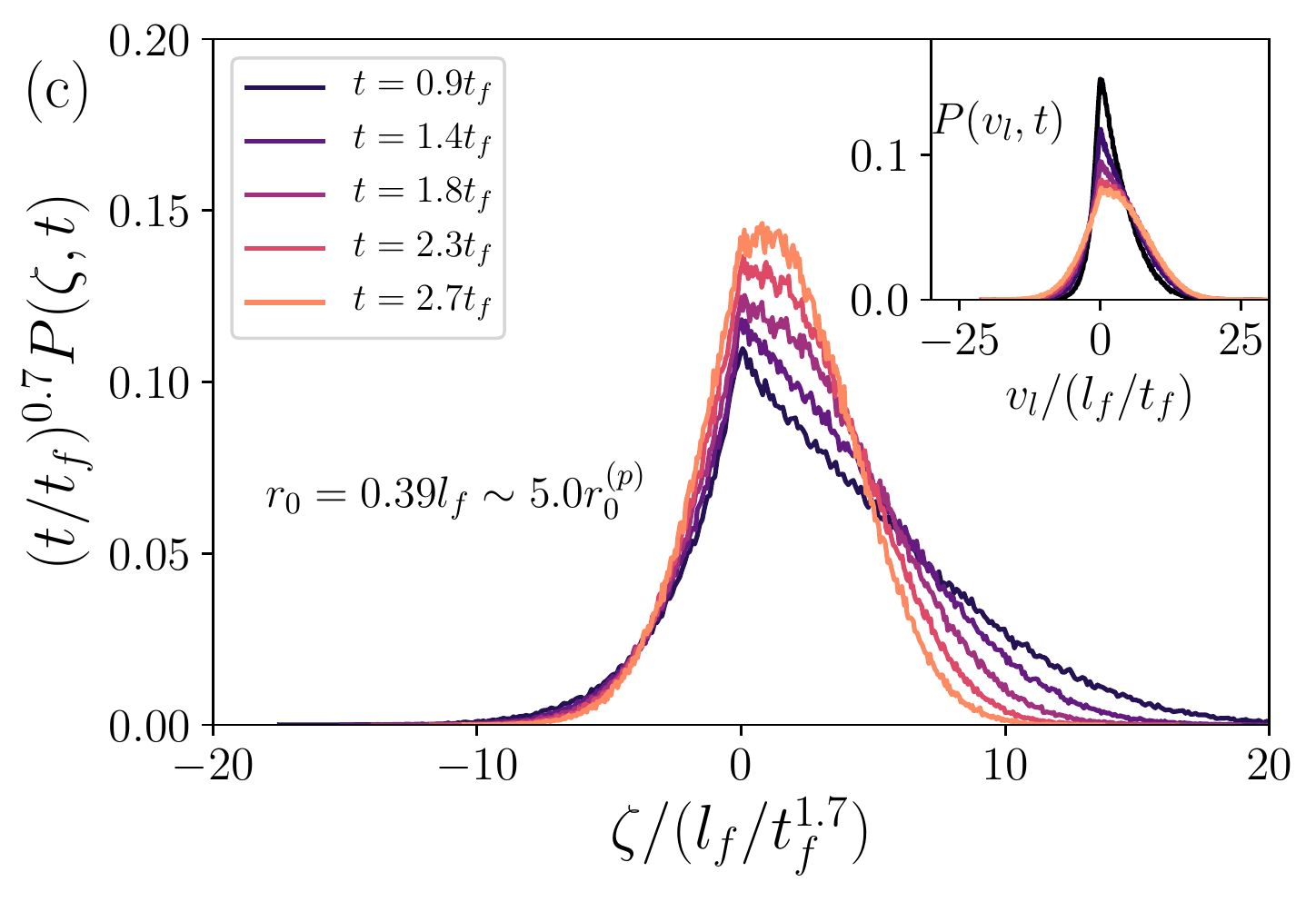}
	\includegraphics[keepaspectratio,scale=0.55]{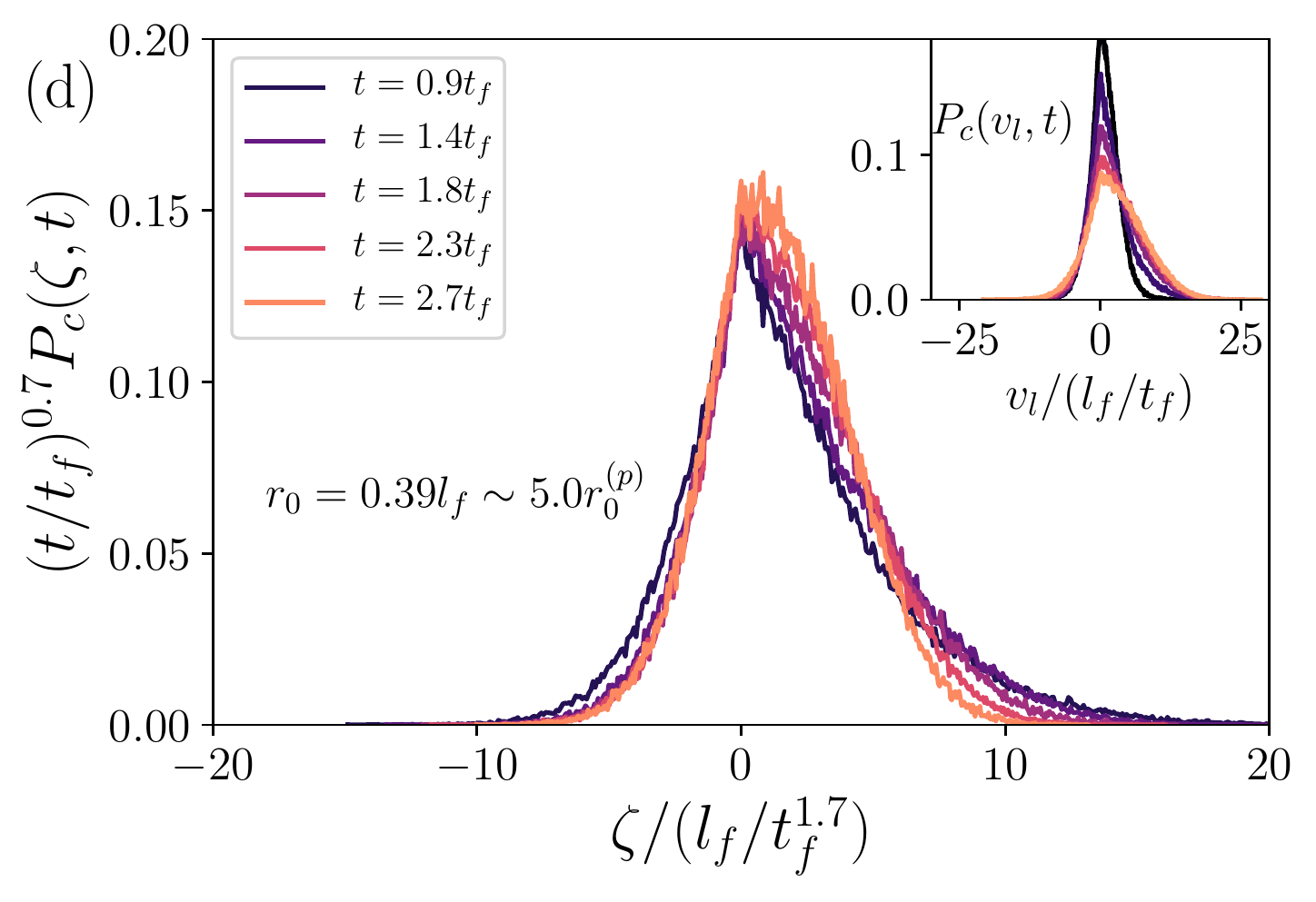}}
	\caption{
		(a) PDFs of 
		the rescaled longitudinal relative velocity,
		$\zeta(t) = v_l(t)/t^{0.7}$ 
		for unconditioned data 
		at different instances $t/t_f = 1.0,1.5, 2.0, 2.5, 3.0$
		for $Re_\alpha = 80$ with hyperviscosity.
		Here, the initial separation corresponds to $r_0=0.48 l_f$ 
		which is different from the proper initial separation.
		It should be noted that $\zeta$ is non-dimensionalized 
		as it is divided by $l_f/t_f^{1.7}$.
		Inset:
		PDFs without re-scaling of the longitudinal 
		relative velocity for the unconditioned data 
		with the initial separation $r_0=0.48 l_f$.
		(b) Same as (a) albeit 
		for the conditionally sampled data.
		(c) Same as (a) albeit for $Re_\alpha = 39$ with normal viscosity
		at different instances $t/t_f = 0.9, 1.4, 1.8, 2.3, 2.7$.
		Here, the initial separation corresponds to $r_0 = 0.39 l_f$.
		(d) Same as (c) albeit for conditionally sampled data.
	}
	\label{fig:pdf_v}
\end{figure}

\begin{figure}[H]
\centerline{
	\includegraphics[keepaspectratio,scale=0.4]{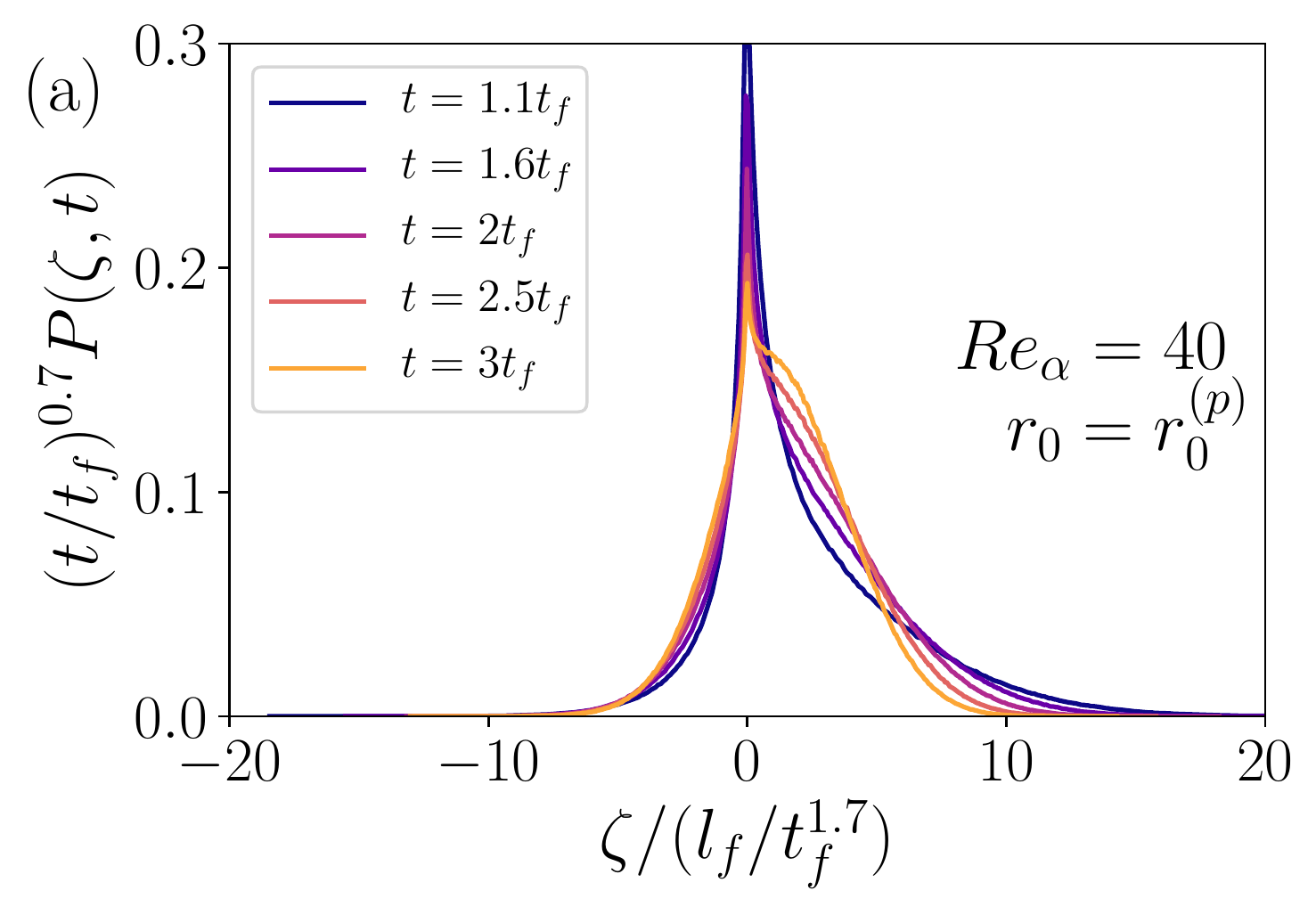}
	\includegraphics[keepaspectratio,scale=0.4]{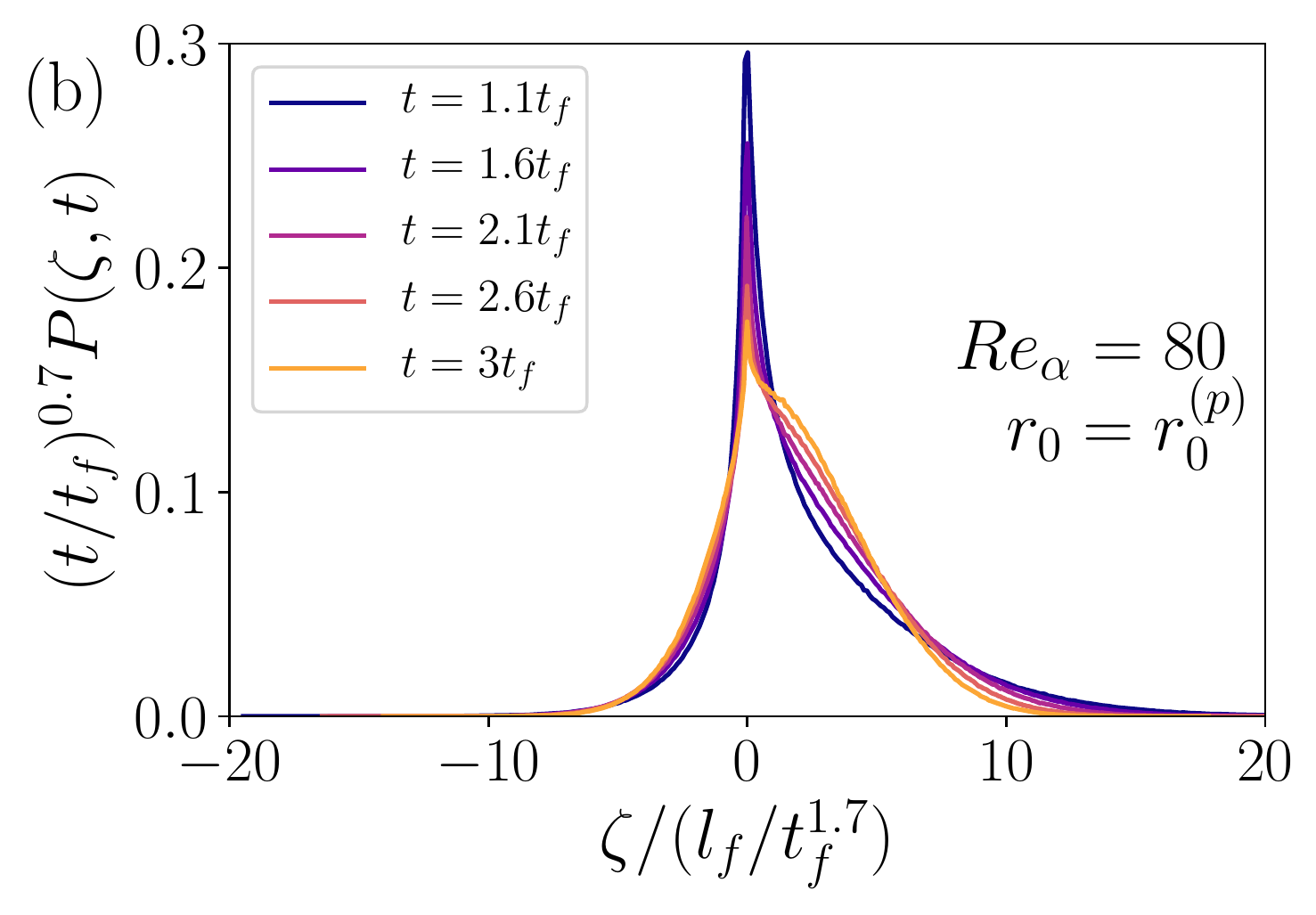}
	\includegraphics[keepaspectratio,scale=0.4]{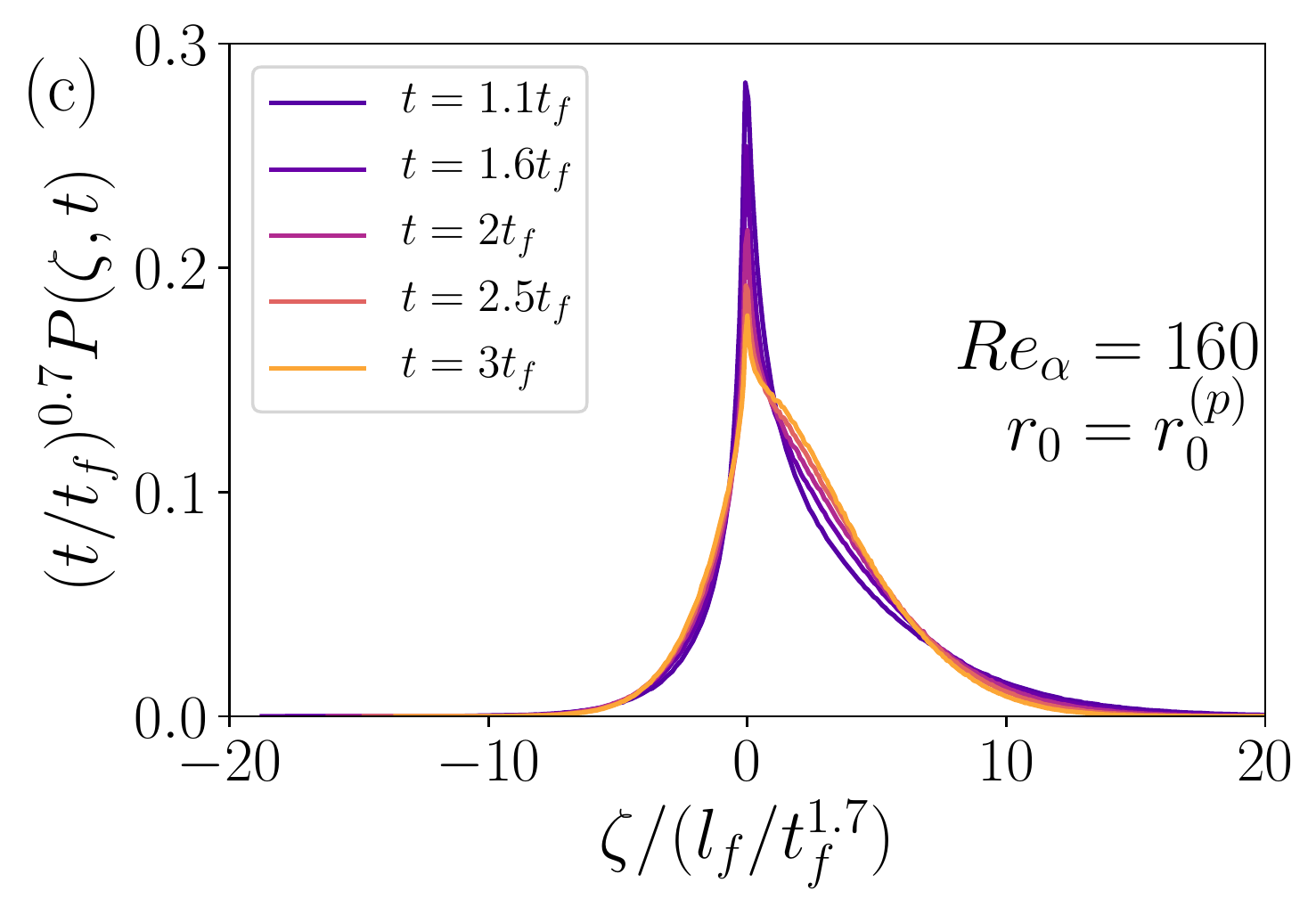}
	}
	\caption{
		Evolution of PDFs of the rescaled longitudinal 
		relative velocity, 
		$\zeta(t) = v_l(t)/t^{0.7}$, of the unconditioned pairs
                starting from the proper initial separation, $\rp$.
		The PDFs are for 
		(a) $Re_\alpha=40$. (b) $Re_\alpha=80$. 
		(c) $Re_\alpha =160$.
	}
	\label{fig:pdf_vp}
\end{figure}

Furthermore, by normalizing with temporal scaling in Eq.~(\ref{eq:v1_scaling}),
the time evolution of the PDF of the conditionally sampled $v_l$
becomes self-similar as shown in Fig.~\ref{fig:pdf_v}(b).
Here, $P_c(A,t)$ corresponds to the conditional PDF for a quantity, $A$.
The collapse among different instances does not appear perfect. 
The collapse around the peak is important 
because the probability in the tails decays faster than the exponential decay
(we compare the degree of the collapse around the peak of the scaled PDF 
to that of the PDF in the inset).
Conversely, the unconditional $v_l$ scaled 
with the same scaling in Eq.~(\ref{eq:v1_scaling})
does not exhibit the self-similar evolution as shown in Fig.~\ref{fig:pdf_v}(a).
Even if we scale the relative velocity with $t^{a/2}$, where the exponent $a$ is measured from 
$\langle v_l(t)^2 \rangle \propto t^{a}$ shown in Fig.~\ref{fig:rv_2}(b) for each $r_0$,
the head parts of the PDFs do not collapse each other as shown in the inset 
of Fig.~\ref{fig:pdf_v}(a).
This implies that the evolution becomes self-similar only for 
conditionally sampled relative velocity 
with scaling relations (\ref{eq:v1_scaling}).
We obtained similar results for the normal viscous case
as shown in Fig.~\ref{fig:pdf_v}(c) and (d).
It should be noted that in the instances plotted in Fig.~\ref{fig:pdf_v},
the conditional separation, $\langle r^2(t) \rangle_c$, is forced to agree
with the Richardson--Obukhov law.
The self-similar evolution of the PDF of $v_l(t)$ also holds for
unconditioned data starting from the proper initial separation as shown
in Fig.~\ref{fig:pdf_vp} for three cases of $Re_\alpha$ with hyperviscosity.

\begin{figure}[H]
\centerline{
	\includegraphics[keepaspectratio,scale=0.55]{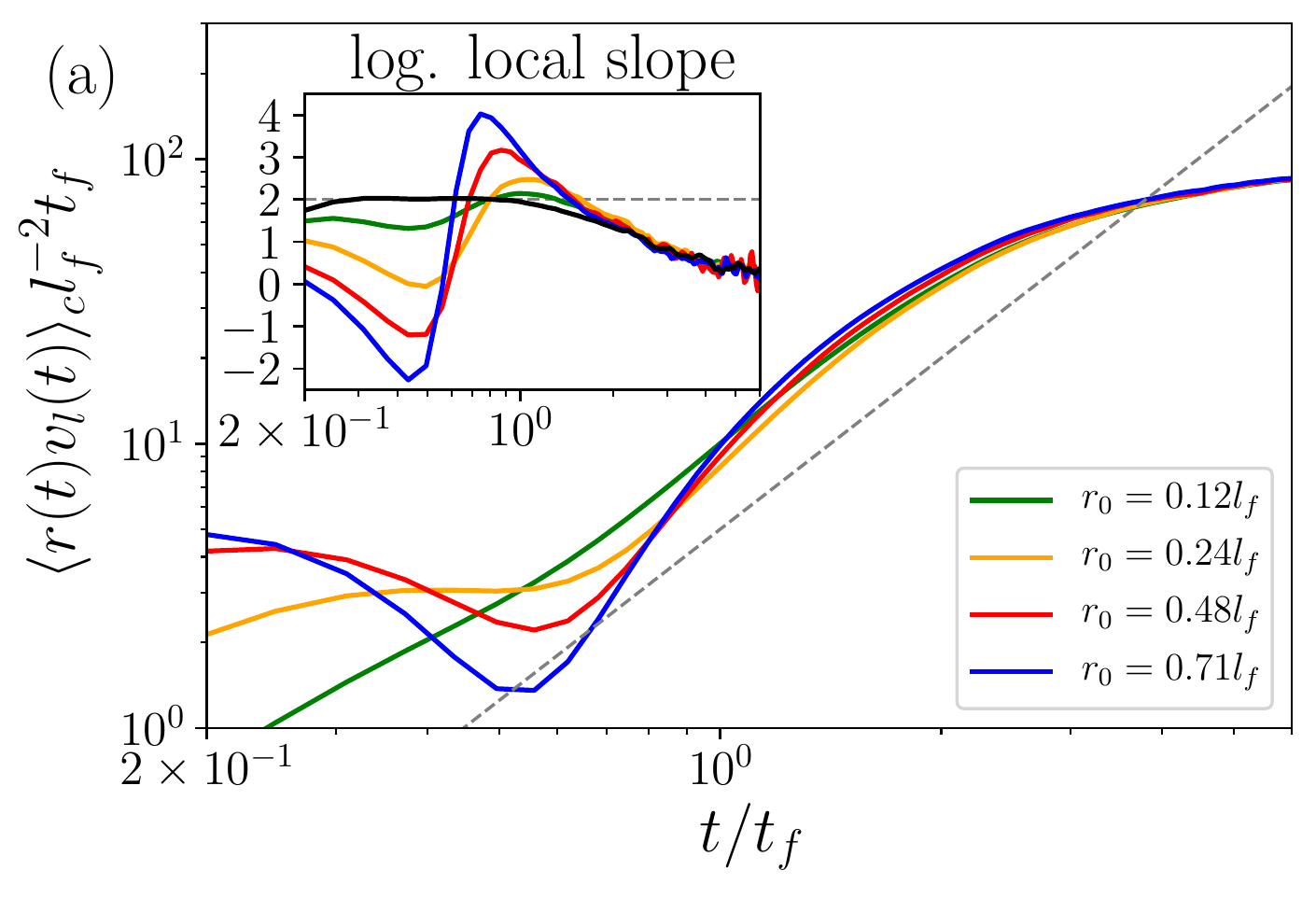}
	\includegraphics[keepaspectratio,scale=0.55]{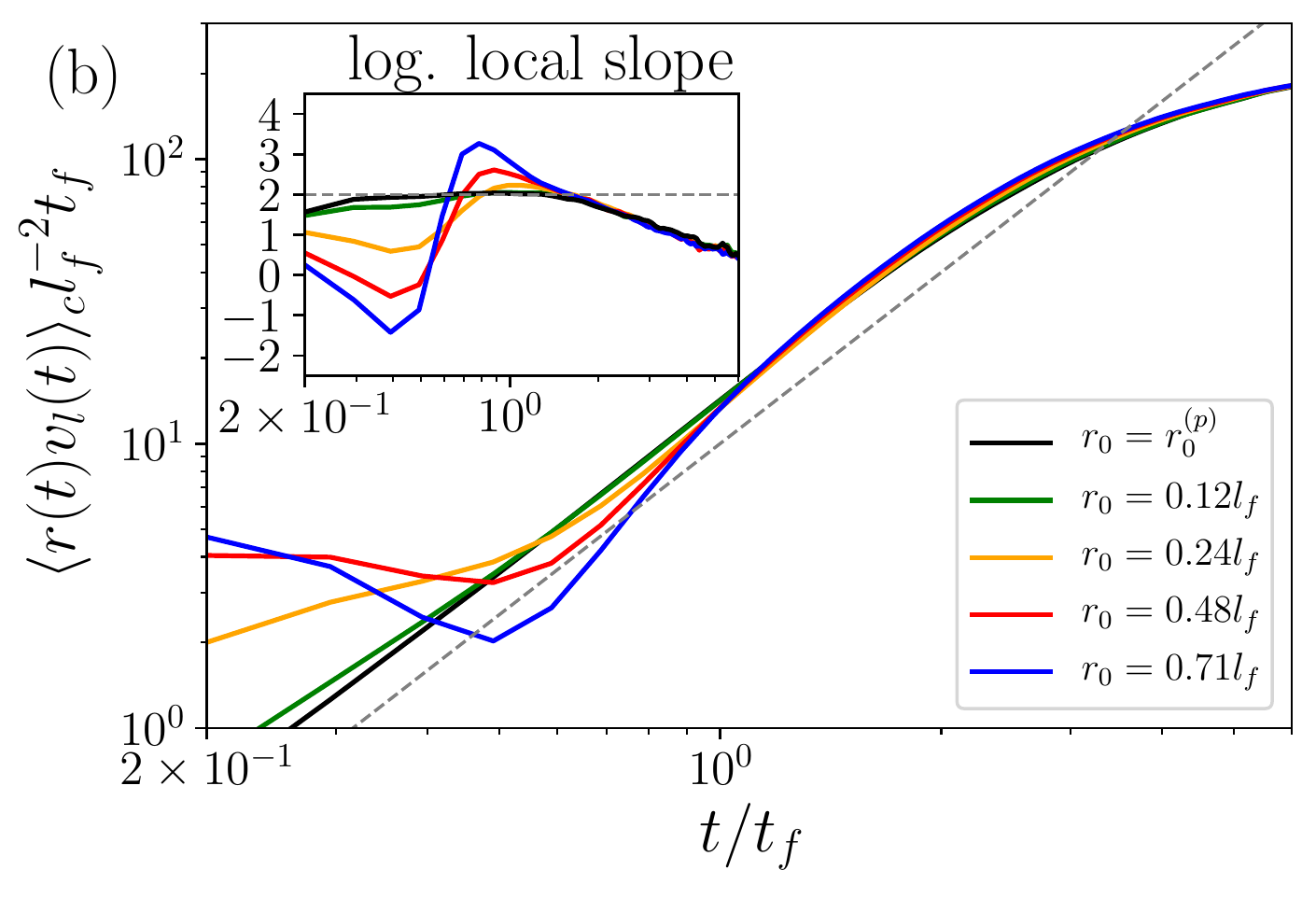}
}
 \caption{
	 (a) Mean separation rate,
	 $\langle r(t) v_l(t) \rangle_c$
	 for conditioned data
	 at various initial separations
	 and for unconditioned data
	 at the proper initial separation, $\rp$
	 for $Re_\alpha = 40$ with hyperviscosity.
	 Dashed line denotes $t^2$ scaling.
	 Inset: The logarithmic local slopes 
	 of the data shown in the outset.
	 Dashed line denotes $t^2$ scaling.
	 (b) Same as (a) but for $Re_\alpha=80$ with hyperviscosity. 
 }
 \label{fig:sep}
\end{figure}

Now, we obtain two evidently incompatible results
via conditional sampling and selecting the proper initial separation.
The second-order moment of the separation, $r(t)$, obeys the K41 scaling 
(although this is enforced).
Conversely, the statistics of relative velocity, $v_l(t)$, deviate
from the K41 scaling although its evolution is self-similar. 
As a soft argument in favor of the compatibility between the two results,
we examine the mean of the product $r(t) v_l(t)$.
It should be noted that it is directly related to the evolution of the mean squared separation
as $d\langle r^2(t) \rangle/dt = 2\langle r(t)v_l(t)\rangle$, and thus it is termed as the separation rate.
As shown in Fig.~\ref{fig:sep}(a), the conditional mean separation rate
obeys $\langle r(t) v_l(t)\rangle_c \propto t^2$ as expected given that we enforced 
the Richardson--Obukhov law. 
The same $t^2$ scaling also holds for the unconditioned mean separation rate 
starting from the proper initial separation (figure not shown).
Evidently, $r(t)$ and $v_l(t)$ are statistically dependent, and this is also evident
from the kinematics.
Hence $\langle r(t) v_l(t) \rangle_c \ne \langle r(t) \rangle_c \langle v_l(t) \rangle_c$.
This indicates that $\langle v_l(t) \rangle_c \sim t^{0.7}$ does not affect
the $t^2$ law of the mean separation rate.
We observe that the mean separation rate 
differs from scaling 
$\langle r(t) \rangle_c \langle v_l(t) \rangle_c 
\sim t^{3/2 + 0.7} = t^{2.2}$ ,
as shown in the insets of Fig.~\ref{fig:sep}(a) and (b), 
if we consider proper initial separation data
as the truly asymptotic data.
Therefore, non-Kolmogorov scaling $\langle v_l(t) \rangle_c \sim t^{0.7}$ is not ruled
out due to the dependence despite the Richardson--Obukhov law $\langle r^2(t) \rangle_c \sim t^3$ 
or, equivalently, the scaling of its time derivative $\langle r(t) v_l(t)\rangle_c \sim t^2$.

\subsection{Quasi-steady state simulation}

The non-K41 scaling of the relative velocity as shown in Fig.~\ref{fig:moment_v}
is not convincing due to the limited scaling range. 
Here, we increase the scaling range by using the quasi-steady state 
of the inverse energy-cascade turbulence \cite{Kraichnan1967}.

Specifically, we solve the Navier--Stokes equation, Eq.~(\ref{eq:NS}),
without the hypodrag term, i.e., $\alpha=0$ by maintaining the other parameters 
as identical to those in Table \ref{tab:parameter} with hyperviscosity ($h = 8$).
With respect to averaging, we generate ten random initial data with flat energy 
spectra extending up to the truncation wavenumber $k_{\max} = (N + 2)/3$ with kinetic energy corresponding to $0.010$.
Over the ten runs, we take the ensemble average.
We perform the simulation with the three resolutions corresponding to $N = 1024, 2048$, and $4096$. 
We use the statistically quasi-steady velocity field obtained in time 
$24.0 \le t \le 26.5$ for advecting the particle pairs. 
In the time window,
the energy spectrum shows the $k^{-5/3}$ scaling extending down to approximately $k = 1$
and the energy grows linearly in time as $\varepsilon t$.
Here, we do not use conditional sampling
and consider only the particle pairs starting 
from the proper initial separation estimated as 
$\rp = 0.60 \times (2\pi/N)$ for each resolution, 
which amounts to $0.145 l_f$.
The value exceeds those of the statistically steady state, 
$\rp = 0.078 l_f$ at $Re_\alpha = 39$ with normal viscosity and,
$\rp  = 0.089 l_f$ at $Re_\alpha =40$,
$\rp  = 0.104 l_f$ at $Re_\alpha =80$,
and $\rp  = 0.122 l_f$ at $Re_\alpha =160$ with hyperviscosity.
This indicates that $\rp$ is affected by the cut-off scale of the inertial range
because small-scale quantities are 
expected to be identical to steady-state simulations.

\begin{figure}[H]
\centerline{
	\includegraphics[keepaspectratio,scale=0.55]{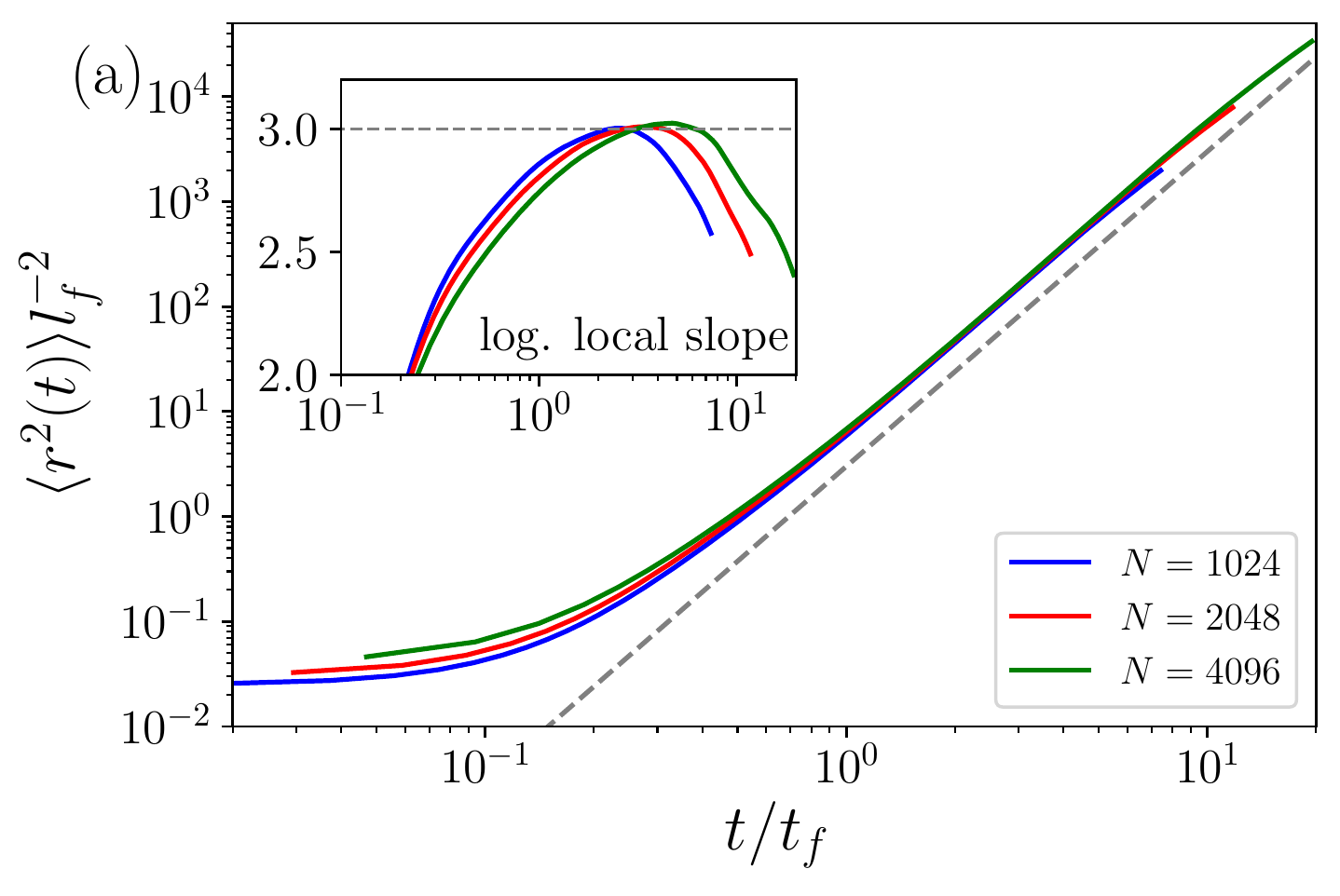}
	\includegraphics[keepaspectratio,scale=0.55]{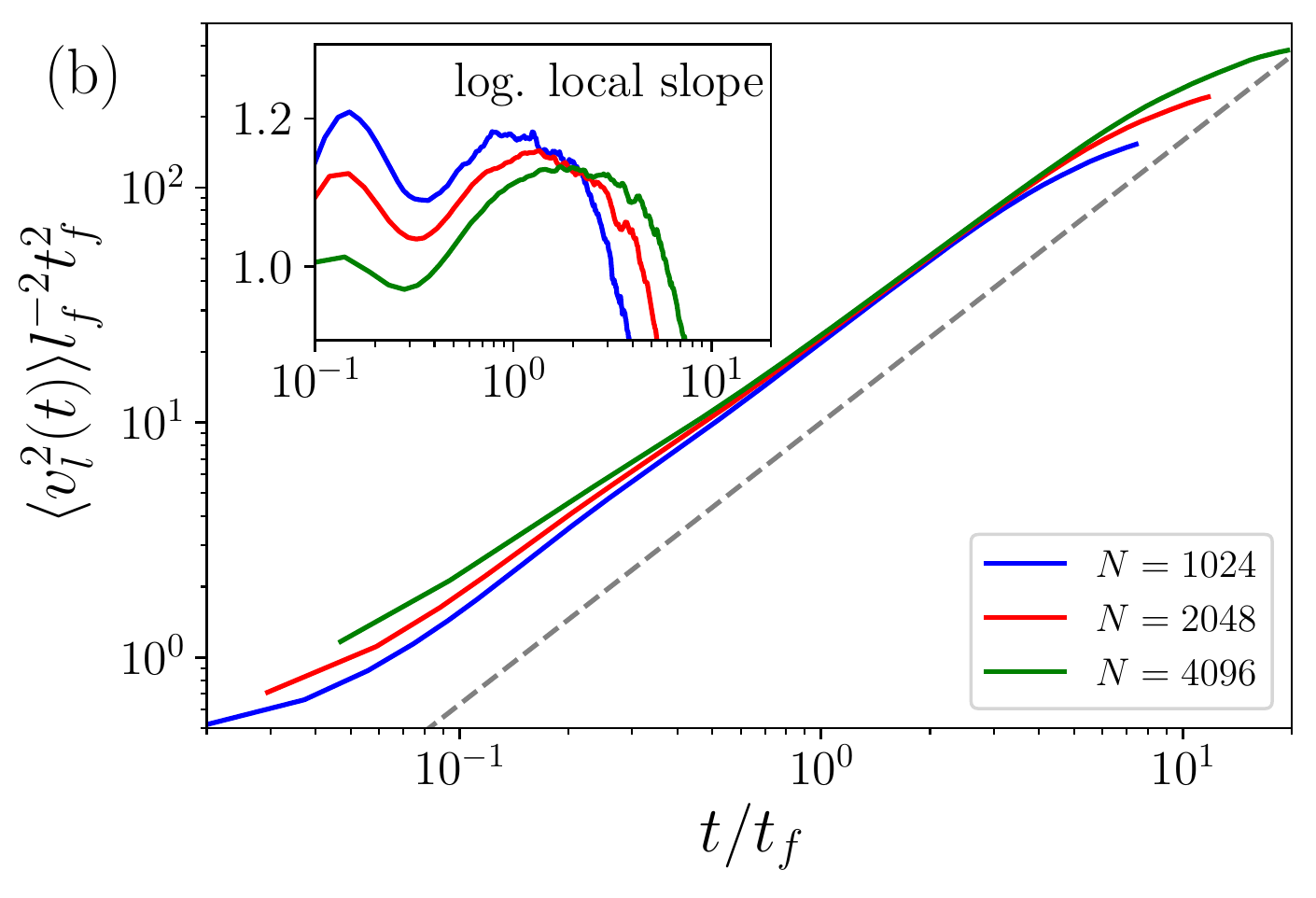}
}
 \caption{
	 (a) Second-order moments of the relative separations,
	 $\langle r^2(t) \rangle$, starting from the proper initial separations
	 in the quasi-steady simulations with resolution $N=1024, 2048$, and $4096$. 
	 Dashed lines denote $t^3$ scaling.
	 Inset:	the logarithmic local slope of 
	 $\langle r^2(t) \rangle$.
	 (b) Same as (a) albeit for the second-order moments of the longitudinal
	 relative velocity, $\langle v_l^2(t) \rangle$.
	 Dashed line denotes $t^{1.2}$ scaling.
	 Inset: the logarithmic local slope of 
	 $\langle v_l^2(t) \rangle$.
 }
 \label{fig:unsteady}
\end{figure}

Figure \ref{fig:unsteady}(a) shows $\langle r^2(t) \rangle$ satisfying
the $t^3$ scaling law for longer duration than statistically steady-state cases.
In Fig.~\ref{fig:unsteady}(b), we present $\langle v_l^2(t) \rangle$ that confirms
the non-K41 power-law scaling observed in the statistically steady-state simulations. 
In more precise terms, from the logarithmic local slope in the inset of Fig.~\ref{fig:unsteady}(b), 
we estimate that the scaling exponent is approximately $1.2$. This is consistent 
with the relation~(\ref{eq:v2_scaling}).
We note that the slopes in the inset of Fig.~\ref{fig:unsteady}(b) 
do not exhibit well-developed plateaus.

To summarize Sec.~\ref{sec:results}, 
we find the non-K41 scaling law of the relative velocity, $v_l(t) \propto t^{0.7}$,
and self-similar evolution of the PDFs of $v_l(t)$ in the two selected ensembles 
of the particle pairs. 
An ensemble corresponds to pairs starting from the proper initial separation $\rp$. 
The other ensemble corresponds to the conditional sampling of the pairs starting from $r_0 > \rp$.
For both ensembles, the Richardson--Obukhov law, $\langle r^2(t) \rangle = g \varepsilon t^3$,
is designed to hold.

\section{Concluding Remarks
\label{sec:discussion}
}

In the 2D inverse energy-cascade turbulence, we developed conditional 
sampling to recover 
the Richardson--Obukhov law by using the relation between 
the exit-time PDF and Richardson PDF.
The conditional squared separation obeys the Richardson--Obukhov law,
$\langle r^2(t) \rangle_c = g \varepsilon t^3$, irrespective of the initial 
separation $r_0$. It is noted that we mainly considered $r_0 > \rp$.
The fraction of the particle pairs remaining in the conditional sampling
increased with increases in $Re_\alpha$. 
This supports our assumption that a bulk of the particle pairs for various 
initial separations at the moderate Reynolds numbers are in agreement with 
the Richardson--Obukhov law. As $Re_\alpha \to \infty$, deviation  in
$\langle r^2(t) \rangle$ from the Richardson--Obukhov law $g \varepsilon t^3$
is likely to vanish. This leads to a conclusion similar to that in
a study of the Richardson--Obukhov law in 3D \cite{D.Buaria2015}.
Furthermore, conditional sampling indicated that the relative velocity 
exhibits a different temporal scaling from the prediction of the K41. 
The results are also obtained for the normal viscous and 
hyperviscous cases.
Therefore, we conclude that the hyperviscosity does not affect the statistical properties
of particle pairs.

Evidently, it is always possible to devise conditional sampling to obtain 
any desired result.
To avoid the pitfall, we showed that the conditional statistics are 
weakly dependent on the parameters, number of the monitored zones $N_Q$, and 
the thresholds of the exit time $\tau$'s. 
An important finding is 
that $N_Q = 1$ is sufficient. This implies that the deviation from 
the Richardson--Obukhov law is caused in the dissipation range and also
by the forcing. 
It implies that a major deviation is not produced later 
in the inertial range.
The latter implication can result from the intermittency-free Eulerian velocity 
field of the 2D inverse energy-cascade turbulence.

However, the implications can overlook the behavior of pairs starting from 
the proper initial separation for which the deviation is negligible.
The results indicated that
the self-similar evolution of the longitudinal relative velocity
is a common feature between the conditionally sampled pairs 
and unconditional pairs staring from $\rp$.
This self-similarity is not observed in the unconditional pairs starting 
from $r_0 > \rp$.
It should be noted that the self-similarity emerges only
with the non-K41 power law of the equal-time relative velocity correlation,
namely the relation~(\ref{eq:v2_scaling}).  
Furthermore, the self-similarity among the PDFs of various instances indicates that the
non-K41 scaling differs from the intermittency observed in the Eulerian velocity 
increments of 3D turbulence. We argued that the non-K41 velocity scaling
is not immediately ruled out by the enforced Richardson--Obukhov law. 

The non-K41 power-law scaling obtained here, $\langle v_l^2(t) \rangle  \propto t^{1.23}$,
exhibits an exponent that differs from the K41 prediction, $\langle v_l^2(t) \rangle  \propto \varepsilon t$.
This can be qualitatively explained 
by the following behavior of the two-time correlation function of 
the Lagrangian relative velocity, $\langle \delta {\bm v}(s_1) \cdot \delta {\bm v}(s_2) \rangle$,
where $\delta {\bm v}(t) = {\bm v}(t|{\bm a}+ {\bm r}_0) - {\bm v}(t|{\bm a})$.
We use DNS data starting from the proper initial separation and plot
the correlation function in the 2D space $(s_1, s_2)$. This type of a plot is presented 
for the 3D case in \cite{ishihara2002}. The two-time correlation is characterized 
by two functional forms as follows: one  along the diagonal line and the other along the line perpendicular 
to the diagonal line. The preliminary study suggests that the two functional forms exhibit 
distinct self-similar functions. Specifically, we speculate that 
the self-similarity of the latter one along the line normal to the diagonal line
leads to the deviation from the K41 scaling of the relative velocity.
Thus, the non-K41 behavior of the velocity can be ascribed to the temporal 
correlation, which is ignored in the K41 argument \cite{Falkovich2001, Ogasawara2006b}.
A future study will detail the two-time correlation.

The results obtained with the enforced Richardson--Obukhov law
lead us to conclude that self-similarity of the relative
velocity with the non-K41 scaling plays an indispensable role in 
the Richardson-Obukhov law of the squared separation. 
The condition is fulfilled for the pairs starting from the proper initial 
separation, $\rp$. An explanation for this is absent.
It can be cautiously stated that quantitative aspects of the proper initial separation 
depend on the forcing because $\rp < l_f$.

We qualitatively discuss the characteristics of the special particle pairs
initially separated by $\rp$ with respect to conditional sampling.
The conditional sampling classifies particle pairs into three groups as follows:
(i) removed particles for $r_0>\rp$, 
(ii) removed particles for $r_0<\rp$, and 
(iii) unremoved particles.
It should be noted that we here include the result of the conditional sampling for $r_0 < \rp$.
We argue that the nature of each group can be different.
For $r_0 > \rp$, the power-law exponent of the unconditional $\langle r^2(t) \rangle$
is lower than the Richardson--Obukhov exponent $3$ as shown in Fig.~\ref{fig:rv_2}(a).
In the conditional sampling, we remove particle pairs in which the exit time per the mean 
is lower than the threshold, $\tau$. Subsequently, the power-law exponent of $\langle r^2(t) \rangle_c$ 
rises to 3. This implies that the removed pairs for $r_0 > \rp$ lower the power-law exponent of $\langle r^2(t) \rangle$.

A physical interpretation can be as follows. 
The removed pairs in the group (i) typically either hardly expand and consequently
stay at around the initial separation or exit from
the inertial range and then behave as standard Brownian particles
while the unremoved particle pairs are still in the inertial range.
Conversely,
for $r_0 < \rp$, the power-law exponent of $\langle r^2(t) \rangle$ is larger
than $3$ as shown in Fig.~\ref{fig:rv_2}(a).
In the conditional sampling, we remove the particle pairs in which the exit time per the mean is 
within the interval, $[\tau_1, \tau_2]$. Subsequently  the power-law exponent of 
$\langle r^2(t) \rangle_c$ decreases to 3.
This implies that the removed pairs for $r_0 < \rp$ increase the power-law exponent of $\langle r^2(t) \rangle$.
A physical interpretation is as follows.
The removed  pairs for $r < \rp$ in the group (ii) typically expand anomalously fast through the inertial range 
while the unremoved particle pairs are still in the inertial range.
The pairs in the group (iii), namely, the unremoved pairs in the conditional sampling regardless of
the initial separation, are typically those that satisfy the Richardson--Obukhov law. 
The results indicated that the fraction of the pairs belonging to the groups (i) and (ii)
significantly depend on the initial separation.
Groups (i) and (ii) are potentially related to the extreme events \cite{Scatamacchia2012,Biferale2014}.
We now return to the proper initial separation.
It is inferred that the effects of the two removed groups on $\langle r^2(t) \rangle$ 
are balanced at the proper initial separation.
Hence, the Richardson--Obukhov law recovers for $\rp$ 
without the conditional sampling 
because contamination from the two groups is cancelled.
Additionally, the cancelling  also supports the dependence of the proper initial separation on the width of
the inertial range mentioned in Sec.IVB, i.e.,  $\rp$ increases with $Re_\alpha$.
The number of particle pairs in the group (i) 
that exit the inertial range relatively fast
and separate based on the $t^2$ law 
decreases inversely with the width of the inertial range, and
the value of $\rp$ should be increased to cancel the anti-effects of groups (i) and (ii)
on the scaling exponent.

We observed the non-Kolmogorov scaling law of 
the Lagrangian velocity.
Evidently, an important question is whether or not the deviation from
the K41 exponent persists when the Reynolds number increases.
The trend shown in Fig.\ref{fig:slope_v}(a) indicates that the deviation persists. 
However,  it is not possible to eliminate the possibility that 
the Kolmogorov scaling law $\langle v_l^2(t) \rangle \propto t$ 
prevails at significantly higher Reynolds numbers. 
To address the question, 
an approach that differs from numerical simulation 
such as Lagrangian two-point closure theory, is preferable.

Our conditional sampling method can be easily adopted to 3D turbulence.
However, the insights gained in 3D should significantly differ from those 
obtained here in the 2D inverse energy-cascade 
turbulence. 
Physics of the 2D energy inverse-cascade turbulence considerably differs
from that of the 3D turbulence although the scaling argument 
using the dissipation rate (i.e., the mean energy flux) leads to the same prediction of scaling exponents of various statistics.
The main difference is that it is necessary to add the forcing at a small scale for the 2D case. This implies that Lagrangian particles in 2D turbulence are 
more directly affected by the forcing than those in 3D turbulence.
A future study will present a detailed analysis of the 3D problem.

\begin{acknowledgments}

Numerical computations in the work were performed
at the Yukawa Institute Computer Facility.
The authors acknowledge support from Grants-in-Aid for Scientific
Research KAKENHI (B) No. 26287023 and KAKENHI (A) No. 19H00641 from JSPS. 
This study was supported by the Research Institute 
for Mathematical Sciences, a Joint Usage/Research Center 
located in Kyoto University.

\end{acknowledgments}

%

\end{document}